\pdfoutput=1
\documentclass[acmsmall,screen,nonacm]{acmart}

\usepackage{booktabs}
\usepackage{multirow}
\usepackage{tikz}
\usepackage{pgfplots}
\pgfplotsset{compat=1.17}
\usepgfplotslibrary{statistics}
\colorlet{palBlue}{blue!60}\colorlet{palOrange}{orange!70}\colorlet{palTeal}{teal!60}\colorlet{palViolet}{violet!55}\colorlet{palGray}{black!35}
\colorlet{mAstra}{palBlue}\colorlet{mStar}{palOrange}\colorlet{mAlpha}{palTeal}\colorlet{mLow}{palViolet}\colorlet{mNone}{palGray}
\colorlet{aContrib}{palBlue}\colorlet{aIxg}{palOrange}\colorlet{aPath}{palTeal}
\colorlet{gMo}{palTeal}\colorlet{gOp}{palBlue}\colorlet{gOx}{palViolet}
\colorlet{sOver}{palBlue}\colorlet{sCorr}{palOrange}\colorlet{sShift}{palTeal}
\pgfplotsset{paperfig/.style={width=0.8\linewidth, height=4.2cm,
  tick label style={font=\scriptsize}, label style={font=\scriptsize}, tick style={black!60}, axis line style={black!60},
  ymajorgrids, grid style={black!12}, enlarge x limits=false,
  legend style={font=\scriptsize, at={(0.5,1.03)}, anchor=south, draw=none, fill=none, legend columns=-1, /tikz/every even column/.append style={column sep=6pt}}},
  paperbars/.style={ybar, area legend},
  paperline/.style={thick, mark size=1.6pt, mark options={solid, draw=black!60, line width=0.4pt}}}
\usetikzlibrary{arrows.meta,positioning,fit,calc,backgrounds,shapes.geometric,decorations.pathreplacing}
\usepackage{enumitem}
\usepackage{algorithm}
\usepackage[noend]{algpseudocode}
\setlist{nosep,leftmargin=1.4em}

\setcopyright{none}
\renewcommand\footnotetextcopyrightpermission[1]{}

\newcommand{\ablAllDIxg}{11.0}
\newcommand{\ablAllDMargin}{31.4}
\newcommand{\ablAllDMaxSeq}{47.0}
\newcommand{\ablAllDMinSeq}{27.9}
\newcommand{\ablAllDOld}{4.4}
\newcommand{\ablAllDOne}{6.5}
\newcommand{\ablAllDPath}{8.1}
\newcommand{\ablAllDSeq}{34.2}
\newcommand{\ablAllSigOld}{4}
\newcommand{\ablAllSigOne}{4}
\newcommand{\ablAllSigSeq}{6}
\newcommand{\attrCellsBest}{5}
\newcommand{\attrCellsSig}{3}
\newcommand{\attrCellsTotal}{6}
\newcommand{\attrMoRuDir}{74.6}
\newcommand{\attrMoRuIxg}{64.0}
\newcommand{\attrMoRuPath}{68.6}
\newcommand{\attrOpDir}{75.9}
\newcommand{\attrOpIxg}{53.6}
\newcommand{\attrOpPath}{54.5}
\newcommand{\attrOxRuDir}{80.5}
\newcommand{\attrOxRuIxg}{82.6}
\newcommand{\attrOxRuPath}{85.6}
\newcommand{\avgAllAstra}{66.7}
\newcommand{\avgAllBest}{48.4}
\newcommand{\avgAllBestName}{STAR}
\newcommand{\avgAllNone}{20.1}
\newcommand{\cellsAllSig}{5}
\newcommand{\cellsAllTotal}{6}
\newcommand{\chAllGainMax}{22.6}
\newcommand{\chAllGainMaxNone}{13.0}
\newcommand{\chAllGainMaxType}{stabilized functions}
\newcommand{\chAllGainMin}{15.6}
\newcommand{\chAllRestNoneMax}{27.0}
\newcommand{\chAllRestNoneMin}{21.1}
\newcommand{\chCellsMaxLoss}{0.8}
\newcommand{\chCellsTotal}{24}
\newcommand{\chCellsWin}{22}
\newcommand{\chNDeprecatedPy}{68}
\newcommand{\chNDeprecatedRu}{27}
\newcommand{\chNImplicitPy}{201}
\newcommand{\chNImplicitRu}{217}
\newcommand{\chNSignaturePy}{133}
\newcommand{\chNSignatureRu}{242}
\newcommand{\chNStabilizedPy}{220}
\newcommand{\chNStabilizedRu}{222}
\newcommand{\gainAllBest}{18.3}
\newcommand{\gainAllCellMax}{31.9}
\newcommand{\gainAllCellMin}{2.1}
\newcommand{\gainAllHi}{19.8}
\newcommand{\gainAllLo}{16.7}
\newcommand{\gainBest}{15.6}
\newcommand{\gainHi}{17.4}
\newcommand{\gainLo}{13.9}
\newcommand{\gainSmBest}{23.5}
\newcommand{\gainSmHi}{26.3}
\newcommand{\gainSmLo}{20.6}
\newcommand{\holmAllNot}{2}
\newcommand{\holmAllSig}{16}
\newcommand{\holmAllTotal}{18}
\newcommand{\keptMo}{6}
\newcommand{\keptOp}{8}
\newcommand{\keptOx}{8}
\newcommand{\layerLastThreeMin}{97.0}
\newcommand{\layerNeverPct}{58}
\newcommand{\layerOpTwentyFour}{55.3}
\newcommand{\layerOxTwentyFour}{1.4}
\newcommand{\layerRepairsMo}{1330}
\newcommand{\layersMo}{24}
\newcommand{\layersOp}{32}
\newcommand{\neuronShareMax}{0.52}
\newcommand{\neuronShareMin}{0.22}
\newcommand{\neuronShareSmMax}{0.95}
\newcommand{\neuronShareSmMin}{0.76}
\newcommand{\ocAllPyAstraTest}{44.7}
\newcommand{\ocAllPyNoneTest}{72.0}
\newcommand{\ocAllRuAstraApi}{2.2}
\newcommand{\ocAllRuAstraCompile}{11.6}
\newcommand{\ocAllRuBestApi}{8.4}
\newcommand{\ocAllRuBestCompile}{26.8}
\newcommand{\ocAllRuNoneApi}{24.8}
\newcommand{\ocAllRuNoneCompile}{44.8}
\newcommand{\ocApiCells}{6}
\newcommand{\ocCompileCells}{3}
\newcommand{\ocOtherMaxAstra}{3.1}
\newcommand{\oneViewAll}{59.3}
\newcommand{\oneViewBestAll}{46.8}
\newcommand{\oneViewBestName}{STAR}
\newcommand{\ovAccMoAstra}{$-$0.69}
\newcommand{\ovAccOpAstra}{+0.45}
\newcommand{\ovAccOxAstra}{+0.81}
\newcommand{\ovBleuMaxOther}{2.4}
\newcommand{\ovBleuMoAlpha}{$-$10.77}
\newcommand{\ovBleuMoAstra}{$-$4.11}
\newcommand{\ovBleuMoStar}{$-$1.66}
\newcommand{\ovBleuOpAlpha}{$-$1.42}
\newcommand{\ovBleuOpAstra}{$-$0.14}
\newcommand{\ovBleuOxAlpha}{$-$4.83}
\newcommand{\ovBleuOxAstra}{$-$2.03}
\newcommand{\ovHeldOut}{32}
\newcommand{\ovRepairs}{132}
\newcommand{\pOxPyAstra}{47.3}
\newcommand{\pOxPyLow}{45.2}
\newcommand{\pOxPyStar}{44.1}
\newcommand{\parMax}{0.058}
\newcommand{\pooledAllLossAlpha}{126}
\newcommand{\pooledAllLossLow}{153}
\newcommand{\pooledAllLossStar}{220}
\newcommand{\pooledAllN}{3990}
\newcommand{\pooledAllPStar}{$p<0.001$}
\newcommand{\pooledAllWinAlpha}{1258}
\newcommand{\pooledAllWinLow}{1403}
\newcommand{\pooledAllWinStar}{961}
\newcommand{\ratioMax}{3.2}
\newcommand{\ratioMin}{1.3}
\newcommand{\relCellsLower}{6}
\newcommand{\relDirMax}{0.24}
\newcommand{\relDirMin}{0.07}
\newcommand{\relMoRuIxg}{0.10}
\newcommand{\relOpPyIxg}{0.55}
\newcommand{\relOpRuPath}{0.35}
\newcommand{\relOthMax}{0.55}
\newcommand{\relOthMin}{0.09}
\newcommand{\relOxPyIxg}{0.30}
\newcommand{\relOxRuPath}{0.19}
\newcommand{\robAllOrigAstra}{64.7}
\newcommand{\robAllOrigLow}{34.4}
\newcommand{\robAllOrigNone}{17.4}
\newcommand{\robAllSpecAlpha}{35.5}
\newcommand{\robAllSpecAstra}{70.9}
\newcommand{\robAllSpecLow}{37.5}
\newcommand{\robAllSpecNone}{13.6}
\newcommand{\robAllSpecStar}{46.1}
\newcommand{\robCellsAheadAll}{6}
\newcommand{\robGapOrigBest}{17.4}
\newcommand{\robGapSpecBest}{22.0}
\newcommand{\robMaxDropAstra}{3.4}
\newcommand{\rqCorrMax}{0.53}
\newcommand{\rqCorrMedOxMax}{0.39}
\newcommand{\rqCorrMedOxMin}{0.37}
\newcommand{\rqCorrMin}{0.37}
\newcommand{\rqCorrSampleMax}{0.65}
\newcommand{\rqCorrSampleMin}{0.23}
\newcommand{\rqCosAbs}{0.009}
\newcommand{\rqCosSampleAbs}{0.05}
\newcommand{\rqKeepMax}{0.90}
\newcommand{\rqKeepMin}{0.83}
\newcommand{\rqOverIqrMax}{0.06}
\newcommand{\rqOverMax}{0.20}
\newcommand{\rqOverMedOxMax}{0.20}
\newcommand{\rqOverMedOxMin}{0.19}
\newcommand{\rqOverMin}{0.15}
\newcommand{\rqOverPctMax}{20}
\newcommand{\rqOverSampleMax}{0.29}
\newcommand{\rqSamples}{1332}
\newcommand{\rqSettings}{6}
\newcommand{\rqShiftMax}{0.17}
\newcommand{\rqShiftMin}{0.10}
\newcommand{\rqShiftSampleMax}{0.26}
\newcommand{\rqShiftSampleMin}{0.05}
\newcommand{\rqTokMax}{293}
\newcommand{\rqTokMin}{85}
\newcommand{\testAllAvgAstra}{65.9}
\newcommand{\testAllBest}{47.8}
\newcommand{\testAllBestName}{STAR}
\newcommand{\timeAllAlpha}{5.4}
\newcommand{\timeAllAstra}{2.7}
\newcommand{\timeAllCellMaxAstra}{4.1}
\newcommand{\timeAllCellMinAstra}{1.6}
\newcommand{\timeAllCellsFasterStar}{6}
\newcommand{\timeAllLow}{1.3}
\newcommand{\timeAllRatioAlpha}{2.0}
\newcommand{\timeAllRatioStar}{2.2}
\newcommand{\timeAllStar}{5.8}

\newcommand{\astra}{\textsc{Astra}}
\newcommand{\semcon}{contrastive semantics}
\newcommand{\Semcon}{Contrastive semantics}
\newcommand{\RQ}[2]{\textbf{RQ#1 (#2):}}
\newcommand{\rqtext}[1]{RQ#1}
\newcommand{\vect}[1]{\mathbf{#1}}
\newcommand{\answer}[2]{\par\noindent\textbf{Answer to RQ#1.} #2\par}
\newcommand{\ixg}{Input$\times$\allowbreak Gradient}
\newcommand{\numViews}{four}
\newcommand{\tabsmall}{\footnotesize\setlength{\tabcolsep}{3.6pt}\renewcommand{\arraystretch}{0.95}}

\begin{document}

\title{Mind the Gaps: From Failure Attribution to Closed-Form Repair of Code Language Models}

\author{Jian Gu}
\affiliation{\institution{Monash University}\country{Australia}}
\email{jian.gu@monash.edu}

\author{Hongyu Zhang}
\affiliation{\institution{Chongqing University}\country{China}}
\email{hyzhang@cqu.edu.cn}

\author{Chunyang Chen}
\affiliation{\institution{Technical University of Munich}\country{Germany}}
\email{chun-yang.chen@tum.de}

\author{Aldeida Aleti}
\affiliation{\institution{Monash University}\country{Australia}}
\email{aldeida.aleti@monash.edu}

\begin{abstract}
Code language models have become persistent components of software development and must be maintained like the software around them. When a library evolves, a model keeps writing the interface that it saw during training, and the defect recurs in every tool built on the model. Repairing the model itself, instead of adding retrieval or retraining, lets one correction reach all downstream uses. Existing repair methods share one pipeline: they attribute a failure to neurons, select the highest-ranked ones, and apply a generic update. This pipeline assumes that the attributed neurons are the ones to patch and that a generic update fits every failure, and neither assumption has been examined. We examine both on executable API evolution tasks in Python and Rust with three code models and identify two gaps. The targeting gap separates the neurons that failure attribution targets from the neurons that can carry a patch: their top sets, the highest-ranked neurons within the repair budget, have a Jaccard overlap of only \rqOverMin{} to \rqOverMax{}. The tailoring gap separates a generic update from a patch built for the failure: the patches that different carrier neurons need are nearly orthogonal. The two gaps are coupled, since one round of patching already changes the targeted top set. To address both gaps, we propose \astra{}. It targets neurons by \semcon{}, an attribution that scores a neuron by its contribution to the logit contrast between the target token and the produced token. It then tailors the patch by solving one small linear system in closed form, which corrects all failing tokens of a sample jointly and needs neither an optimizer nor a backward pass. \Semcon{} selects significantly better carrier neurons than gradient-based attributions in \attrCellsSig{} of \cellsAllTotal{} settings and comparable ones in the others, and its patches change these neurons \ratioMin{} to \ratioMax{} times less in relative norm and lie in late layers. On average, \astra{} reaches \avgAllAstra{} percent Pass@1, against \avgAllBest{} percent for the best of AlphaEdit, STAR and low-rank adaptation, and repairs a sample in \timeAllAstra{} seconds. It is the best method in all \cellsAllTotal{} settings and for every type of API change, and this advantage persists under an unseen phrasing of the test prompt. Its side effects on unrelated code are small on the large models and larger on the small one, where the same budget changes a larger share of neurons.

\end{abstract}

\maketitle

\section{Introduction}
\label{sec:intro}

Code language models have become persistent components of software development, and they must be maintained like the software around them. A model learns one snapshot of the libraries for which it writes code, whereas libraries deprecate functions, add or remove parameters, change behavior between releases and stabilize unstable interfaces~\cite{robbes2012deprecation,liang2025rustevo}. As a result, the model keeps emitting the interface that it saw during training~\cite{wang2024llmsmeetlibevo,liu2024codeupdatearena}. When a release removes a parameter from a function, for example, a model trained before that release continues to pass it, and the generated code fails against the current library.

In this paper, we study the repair of the model itself. Unlike retrieval or broad retraining, a single repair benefits every downstream application. The task is to change the parameters of the model, given one description of the new interface and one reference solution, so that the model writes code for the current interface when the task is posed without that description. Two families of methods do so. Model editing, as in AlphaEdit~\cite{fang2025alphaedit}, rewrites the parameters that store a piece of knowledge through a closed-form or learned update~\cite{meng2022rome,meng2023memit}, whereas semantic analysis, as in STAR~\cite{gu2025star}, attributes a failure to internal units through their contribution to the output and updates only those units~\cite{gu2023neuronpatching}. Both families inherit the pipeline of fault localization and program repair: they attribute a failure to components, select the highest-ranked ones, and update them.

However, this pipeline rests on two assumptions that, to our knowledge, have not been examined. We find that neither holds on two executable benchmarks of API evolution, PyEvo for Python and RustEvo for Rust, with three open code models. The first assumption is that the neurons to which a failure is attributed are the neurons that should be patched. We call the gap between the neurons that failure attribution targets and the neurons that can carry the patch the \emph{targeting gap}. The \emph{top set} of a ranking is its $K$ highest-ranked neurons, where $K$ is the neuron budget of the repair, and the overlap of two top sets is their Jaccard index. In our measurements, the two rankings correlate only moderately, and the overlap of their top sets is only \rqOverMin{} to \rqOverMax{}. The second assumption is that a generic update rule, which moves the selected neurons along a shared direction or treats the failing tokens one by one, serves every failure. We call the shortfall of such a rule against a patch built for the failure as a whole the \emph{tailoring gap}. Indeed, the patches that the neurons need are nearly orthogonal (cosine of at most \rqCosAbs{}), so no shared direction fits them.

Our key insight is that a failure should be attributed in the coordinates in which it is patched, and patched for all failing tokens at once. A patch changes the write vectors of feed-forward neurons, the vectors that these neurons add to the hidden state whenever they are active. We call a neuron whose write vector is changed a \emph{carrier neuron}, because it carries the patch into the model. We introduce \semcon{}, which scores a neuron by its contribution to the logit contrast between the target token and the produced token. This score is the product of the two quantities through which such a change acts, the activation of the neuron and the direction that the output reads, so a neuron with a high score needs only a small change. On this basis, we propose \astra{}, Analytic Semantic Tailoring for Repair by Attribution. Its attribution is semantic, since it reads neurons in vocabulary-defined semantic coordinates, and its patch is analytic, since it has a closed form. \astra{} repairs a sample in four steps. First, it locates the failing tokens of the reference answer by observing the model under teacher forcing on four renderings of the repair prompt, which we call views. Second, it targets the carrier neurons, which it selects once by \semcon{}. Third, it tailors a joint patch that corrects all failing tokens of all views by solving one small linear system. Fourth, because the patch is exact only at first order, it re-linearizes the repaired model by repeating the observation and the solve for at most ten rounds.

We evaluate \astra{} on two benchmarks and three code models: a small model, MiniCPM5-1B, and two large models, OpenCoder-8B and OxCoder-9B.\footnote{\url{https://huggingface.co/OrionLLM/OxCoder-9B}} Carriers selected by \semcon{} repair significantly better than those of two gradient-based attributions in \attrCellsSig{} of the \cellsAllTotal{} settings and comparably in the others. On average, \astra{} reaches \avgAllAstra{} percent Pass@1, against \avgAllBest{} percent for the best baseline, \avgAllBestName{}, and \avgAllNone{} percent without repair, and repairs a sample in \timeAllAstra{} seconds, \timeAllRatioStar{} times faster than \avgAllBestName{}. It is the best method in all \cellsAllTotal{} settings, significantly so in \cellsAllSig{} of them, and for every type of API change. This advantage persists under an unseen phrasing of the test prompt. Its side effects on unrelated code are small on the large models and larger on the small one.

In summary, this paper makes the following contributions.
\begin{itemize}
\item We identify and measure two gaps in neuron-level repair. The top set of failure attribution and that of the carriers of a patch overlap by only \rqOverMin{} to \rqOverMax{} (targeting gap), and the patches of different carriers are nearly orthogonal (tailoring gap). The gaps are coupled, since one round of patching lowers the overlap of the targeted top set with itself to \rqKeepMin{} to \rqKeepMax{}.
\item We show how to address the targeting gap. \Semcon{}, which is forward-only and scores a neuron in the coordinates of the patch, selects carrier neurons that repair significantly better than those of gradient-based attributions in \attrCellsSig{} of \cellsAllTotal{} settings and comparably in the others. The patch changes these neurons \ratioMin{} to \ratioMax{} times less in relative norm and lies in late layers.
\item We propose \astra{}, which addresses the tailoring gap with a closed-form patch that corrects all failing tokens of a sample jointly. It reaches the highest Pass@1 in all \cellsAllTotal{} settings of three models and two benchmarks at \timeAllAstra{} seconds per repair, and an ablation shows that the joint solve matters most. We also report its side effects and robustness.
\end{itemize}

\section{Related Work}
\label{sec:background}

We study language model repair, the targeted modification of a trained model that corrects a specified failure while preserving unrelated behavior. It differs from classical factual editing in two respects: its target is an executable behavior, here a program that must pass tests, and its answer is a long sequence in which many token positions must be corrected at once. Repair methods change feed-forward neurons, because the feed-forward block of a transformer layer is a key-value memory in which keys detect input patterns and values write vectors to the residual stream~\cite{geva2021ffn}.

\paragraph{Repair of deep neural networks.} Before language models, the repair of deep networks localized suspicious neurons or weights and then searched for a modification~\cite{sohn2019arachne,usman2021nnrepair}, adapted weights or selected inputs for retraining~\cite{zhang2019apricot,ma2018mode}, or computed a provably correct change~\cite{sotoudeh2021provable}. Fault localization in networks follows the same pattern~\cite{eniser2019deepfault,wardat2021deeplocalize}, and both lines mirror spectrum-based fault localization and program repair, in which components are ranked by their association with a failure, a patch is built, and the result is validated on the failing input and on regression inputs~\cite{jones2005tarantula,wong2016survey,gazzola2017survey}. In the same spirit, this paper asks of language models what the evaluation of fault localization asks of programs, namely whether the ranking is useful for the repair that follows.

\paragraph{Interpretability for LM analysis.} Interpretability explains where a model computes a behavior. At the level of single components, knowledge neurons are identified by integrated gradients~\cite{dai2022knowledge,sundararajan2017integrated}, and activation patching estimates the effect of a component by replacing its activation~\cite{zhang2023best_practices_patching}. Circuit analysis then moves to the subgraph of attention heads and feed-forward neurons that implements a behavior. A mathematical framework decomposes a transformer into paths through the residual stream~\cite{elhage2021mathematical}, case studies recover the circuits of induction and of indirect object identification~\cite{olsson2022induction,wang2023ioi}, and automated circuit discovery prunes the computational graph to the edges that a behavior requires~\cite{conmy2023acdc}. Lens tools also read intermediate states in the vocabulary space~\cite{nostalgebraist2020logitlens,belrose2023tunedlens}, and dictionary learning and transcoders expose interpretable features~\cite{dunefsky2024transcoders}. For code models, recent work links neurons to programming concepts and uses them to interpret code generation~\cite{kargaran2025programming_neurons,yin2025neuron_guided_code}. In all these cases, a circuit or an attribution states which components compute a behavior, whereas this paper examines whether these components are the ones that can carry a patch.

\paragraph{Model editing for LM repair.} Locate-then-edit methods identify feed-forward layers that store a fact and write a new association with a closed-form update~\cite{meng2022rome,meng2023memit}. Hypernetwork editors predict updates from one example~\cite{decao2021knowledgeeditor,mitchell2022mend}, and AlphaEdit projects the update into the null space of preserved knowledge~\cite{fang2025alphaedit}. However, studies report that sequential or mass edits degrade general abilities and that edit evaluations can be optimistic~\cite{gu2024harms,meng2025mirage}. More recent work refines the carrier of an edit to individual neurons~\cite{pan2025precise_localization} and extends editing to long-form and graph-structured knowledge~\cite{jiang2025anyedit,zhang2024glame}. For code models, editing has been examined for API knowledge~\cite{li2025modeledit4code}, and mechanistic analysis has been used to repair insecure code generation~\cite{elhusseini2026repairing}. As an alternative, adapters and low-rank adapters train a small set of added weights~\cite{hu2022lora}. Like MEMIT and AlphaEdit, \astra{} writes a closed-form minimum-norm change into feed-forward output projections. It differs in two respects: it targets a logit contrast at every failing token of a long answer instead of one residual at a single position, and it chooses layers and neurons for each sample instead of a fixed window of layers.

\paragraph{Semantics for LM analysis and repair.} Requirements on a model are stated on tokens, whereas the model realizes them in activations, residual states and parameters, and semantics is the operational medium that connects the two. For analysis, feed-forward layers promote concepts in the vocabulary space~\cite{geva2022promoting}, and vocabulary-defined semantics gives every latent state a meaning through the output vocabulary~\cite{gu2024vds}. On this basis, the transition of latent states across layers shows where semantically consequential computation occurs~\cite{gu2025layerfreezing}, a semantic reference frame describes the dynamics of the residual stream~\cite{gu2026semrf}, and latent semantic alignment relates the representations of models of different scale~\cite{gu2025crossscale}. For repair, Neuron Patching and STAR attribute a failure to feed-forward neurons through their semantic effect on the output and derive the patch from the latent directions of the target and the produced token~\cite{gu2023neuronpatching,gu2025star}. Pseudo-inverse tying gives the input embedding and the output projection a consistent token interface for later updates~\cite{gu2026pseudoinverse}. \astra{} follows this line and uses one semantic coordinate system, the logit contrast between the target and the produced token, for locating a failure, targeting its carriers and tailoring its patch.

\paragraph{API evolution and other applications.} Software libraries evolve through deprecated interfaces, changed signatures, changed behavior and stabilized functionality~\cite{robbes2012deprecation,liang2025rustevo}, whereas a code model learns one snapshot. Accordingly, benchmarks examine how models handle such changes~\cite{liu2024codeupdatearena,wu2024versicode,liang2025rustevo}, and studies show that models use deprecated interfaces and hallucinate APIs~\cite{wang2024llmsmeetlibevo,liu2024hallucination_code}. Prompt-level remedies retrieve documentation into the context~\cite{lewis2020rag,zhang2023repocoder}, and training-level remedies continue pretraining or fine-tune on updated code~\cite{liu2024codeupdatearena}. Two further scenarios need the same capabilities: modular reuse decomposes networks into replaceable modules, patches weak models with modules of stronger ones and re-engineers models for a target task~\cite{pan2020decomposing_dnn,qi2022patching,qi2023reusing,qi2024modularizing,bi2025nemo,ngo2026dnn_modularization}, and efficient AI coding lowers the cost of code models through shorter code representations, token-saving formats and compressed contexts~\cite{sun2024ai_coders_grammar,pan2026hidden_cost_readability,sun2025token_sugar,shi2025longcodezip}. All three depend on knowing which parts of a model matter for a behavior and how to change them safely.

\section{\astra: Analytic Semantic Tailoring for Repair by Attribution}
\label{sec:approach}

\begin{figure}[htb]
\centering
\begin{tikzpicture}[
  font=\scriptsize, >={Stealth[length=1.7mm]},
  panel/.style={draw=#1!70!black!55, rounded corners=3pt, fill=#1!9, minimum height=4.3cm, anchor=north west},
  hdr/.style={font=\scriptsize\bfseries, anchor=west, inner sep=0pt},
  note/.style={font=\tiny, text=black!75, align=left, inner sep=1pt},
  cnote/.style={font=\tiny, text=black!75, align=center, inner sep=1pt},
  tok/.style={draw=black!50, fill=white, minimum height=0.30cm, inner sep=1.4pt, font=\ttfamily\tiny},
  bad/.style={tok, fill=red!20, draw=red!65!black},
  tag/.style={draw=#1!70!black!60, fill=white, rounded corners=3.5pt, font=\tiny, inner xsep=4pt, inner ysep=1.5pt, text=#1!60!black},
  flow/.style={->, semithick, black!70},
  wide/.style={->, line width=1.1pt, black!60},
  sel/.style={fill=palBlue, draw=black!60},
  cell/.style={fill=white, draw=black!35},
  pics/lm/.style={code={
    \foreach \a/\ya in {1/0.30,2/0,3/-0.30}{\foreach \b/\yb in {1/0.39,2/0.13,3/-0.13,4/-0.39}{
      \draw[black!30, line width=0.3pt] (-0.34,\ya) -- (0,\yb); \draw[black!30, line width=0.3pt] (0,\yb) -- (0.34,\ya);}}
    \foreach \ya/\cl in {0.30/palTeal,0/palOrange,-0.30/palTeal}{\fill[\cl!85!black] (-0.34,\ya) circle (0.055); \fill[\cl!85!black] (0.34,\ya) circle (0.055);}
    \foreach \yb in {0.39,0.13,-0.13,-0.39}{\fill[palBlue!85!black] (0,\yb) circle (0.055);}
  }},
  pics/sheet/.style={code={
    \draw[black!55, fill=white] (0,0) rectangle (0.62,0.78);
    \foreach \y in {0.62,0.50,0.38,0.26}{\draw[black!35, line width=0.5pt] (0.09,\y) -- (0.53,\y);}
    \draw[black!35, line width=0.5pt] (0.09,0.14) -- (0.36,0.14);
  }},
  pics/foldsheet/.style={code={
    \draw[black!55, fill=white] (0,0) -- (0.44,0) -- (0.62,0.18) -- (0.62,0.78) -- (0,0.78) -- cycle;
    \draw[black!55, fill=black!12] (0.44,0) -- (0.44,0.18) -- (0.62,0.18) -- cycle;
    \foreach \y in {0.62,0.50,0.38,0.26}{\draw[black!35, line width=0.5pt] (0.09,\y) -- (0.53,\y);}
    \draw[black!35, line width=0.5pt] (0.09,0.14) -- (0.32,0.14);
  }},
  pics/bug/.style={code={
    \foreach \x/\y in {-0.115/0.05,-0.125/-0.01,-0.105/-0.07,0.115/0.05,0.125/-0.01,0.105/-0.07}{\draw[black!80, line width=0.35pt] (0,-0.01) -- (\x,\y);}
    \fill[black!85] (0,0.075) circle (0.045);
    \draw[black!85, line width=0.3pt] (-0.02,0.11) -- (-0.05,0.15); \draw[black!85, line width=0.3pt] (0.02,0.11) -- (0.05,0.15);
    \draw[black!80, line width=0.3pt, fill=red!80!black] (0,-0.015) ellipse (0.085 and 0.095);
    \draw[black!80, line width=0.3pt] (0,0.08) -- (0,-0.11);
    \foreach \x/\y in {-0.04/0.025,0.04/0.025,-0.04/-0.045,0.04/-0.045}{\fill[black!85] (\x,\y) circle (0.016);}
  }},
  pics/bandage/.style={code={
    \draw[black!50, fill=orange!35!yellow!45, rounded corners=1.6pt] (-0.27,-0.085) rectangle (0.27,0.085);
    \draw[black!40, fill=orange!12!white] (-0.085,-0.085) rectangle (0.085,0.085);
    \foreach \x in {-0.19,-0.14,0.14,0.19}{\fill[black!45] (\x,0.03) circle (0.011); \fill[black!45] (\x,-0.03) circle (0.011);}
  }},
]
\def\mid{-2.30}%
\node[hdr] at (0,-0.24) {Input};
\pic[scale=1.15] at (0.73,-1.04) {lm};
\pic[scale=1.15, rotate=-20] at (0.73+0.391,-1.04+0.345) {bug};
\node[cnote, anchor=north] at (0.73,-1.56) {language model};
\node[draw=black!55, fill=white, rounded corners=1pt, minimum height=0.34cm, inner sep=2pt, font=\ttfamily\tiny, anchor=north] (ans) at (0.73,-2.12) {wait\_for(f, t)};
\node[cnote, anchor=north] at ($(ans.south)+(0,-0.03)$) {reference answer};
\foreach \k in {3,2,1}{\pic at (0.33+\k*0.06,-4.00+\k*0.05) {sheet};}
\pic at (0.33,-4.00) {foldsheet};
\node[cnote, anchor=north] at (0.73,-4.03) {prompt in views};
\node[panel=palTeal, minimum width=2.82cm] (P1) at (2.04,0) {};
\node[hdr] at ($(P1.north west)+(0.12,-0.24)$) {Locate Failing Tokens};
\node[note, anchor=north west] at ($(P1.north west)+(0.10,-0.46)$) {observe by teacher forcing};
\foreach \i/\t/\s/\w in {0/wait\_for/tok/0.86,1/(/tok/0.26,2/f/tok/0.26,3/{,}/tok/0.26,4/t/tok/0.26,5/)/bad/0.26}{
  \pgfmathsetmacro{\x}{ifthenelse(\i==0,0.14,1.00+0.30*(\i-1)+0.06)}
  \node[\s, anchor=west, minimum width=\w cm] (t\i) at ($(P1.north west)+(\x,-1.08)$) {\t};}
\node[font=\ttfamily\tiny, text=red!65!black, anchor=north east, inner sep=1pt] at ($(t5.south east)+(0.04,-0.03)$) {{,} loop=};
\node[note, anchor=north west, text=red!65!black] at ($(P1.north west)+(0.10,-1.30)$) {produced:};
\draw[flow, black!55] ($(t5.south)+(0,-0.28)$) -- ++(0,-0.22);
\begin{scope}[shift={($(P1.north west)+(0.22,-3.30)$)}]
  \node[note, anchor=south west] at (-0.08,1.18) {logits at the failing token};
  \draw[black!50] (-0.05,0) -- (2.40,0);
  \fill[red!55] (0.08,0) rectangle +(0.26,0.76);
  \fill[green!50!black!60] (0.46,0) rectangle +(0.26,0.38);
  \node[font=\ttfamily\tiny, anchor=base, inner sep=0pt] at (0.21,-0.165) {,};
  \node[font=\ttfamily\tiny, anchor=base, inner sep=0pt] at (0.59,-0.19) {)};
  \draw[densely dotted, black!60] (0.34,0.76) -- (1.35,0.76);
  \draw[densely dotted, black!60] (0.08,1.08) -- (1.74,1.08);
  \draw[densely dotted, black!60] (0.72,0.38) -- (1.74,0.38);
  \draw[<->, black!60] (1.25,0.76) -- (1.25,1.08);
  \node[font=\tiny, anchor=east, inner sep=1.5pt] at (1.16,0.92) {margin};
  \draw[<->, black!75] (1.66,0.38) -- (1.66,1.08);
  \node[font=\tiny, anchor=west, inner sep=1.5pt] at (1.70,0.73) {deficit};
\end{scope}
\draw[bad] ($(P1.south west)+(0.16,0.44)$) rectangle +(0.24,0.18); \node[note, anchor=west] at ($(P1.south west)+(0.44,0.53)$) {failing token};
\draw[tok] ($(P1.south west)+(0.16,0.15)$) rectangle +(0.24,0.18); \node[note, anchor=west] at ($(P1.south west)+(0.44,0.24)$) {passing token};
\node[panel=palBlue, minimum width=2.9cm] (P2) at ($(P1.north east)+(0.50,0)$) {};
\node[hdr] at ($(P2.north west)+(0.12,-0.24)$) {Target Carrier Neurons};
\node[note, anchor=north west] at ($(P2.north west)+(0.10,-0.46)$) {filter by \semcon{}};
\foreach \r in {0,...,7}{
  \foreach \c in {0,...,7}{\draw[cell] ($(P2.south west)+(0.56+\c*0.19,0.84+\r*0.29)$) rectangle +(0.14,0.20);}
}
\foreach \r in {5,7}{\draw[palBlue!80!black, thick, rounded corners=1pt] ($(P2.south west)+(0.52,0.80+\r*0.29)$) rectangle +(1.55,0.28);}
\foreach \r/\c in {7/1,7/3,7/4,7/6,5/0,5/2,5/5,5/7}{\draw[sel] ($(P2.south west)+(0.56+\c*0.19,0.84+\r*0.29)$) rectangle +(0.14,0.20);}
\foreach \r/\w in {0/0.03,1/0.02,2/0.05,3/0.04,4/0.10,5/0.36,6/0.15,7/0.48}{\fill[palBlue!50] ($(P2.south west)+(2.18,0.86+\r*0.29)$) rectangle +(\w,0.16);}
\node[note, rotate=90, anchor=south] at ($(P2.south west)+(0.38,2.00)$) {layers};
\node[note, anchor=south west] at ($(P2.south west)+(2.12,3.13)$) {score};
\node[note, anchor=east] at ($(P2.south west)+(0.54,2.97)$) {$L$};
\node[note, anchor=east] at ($(P2.south west)+(0.54,0.94)$) {$1$};
\draw[palBlue!80!black, thick, rounded corners=1pt] ($(P2.south west)+(0.14,0.43)$) rectangle +(0.34,0.20); \node[note, anchor=west] at ($(P2.south west)+(0.52,0.53)$) {selected layer};
\draw[sel] ($(P2.south west)+(0.24,0.15)$) rectangle +(0.14,0.19); \node[note, anchor=west] at ($(P2.south west)+(0.52,0.24)$) {carrier neuron};
\node[panel=palOrange, minimum width=2.82cm] (P3) at ($(P2.north east)+(0.50,0)$) {};
\node[hdr] at ($(P3.north west)+(0.12,-0.24)$) {Tailor Joint Patches};
\node[note, anchor=north west] at ($(P3.north west)+(0.10,-0.46)$) {closed-form linear solve};
\foreach \r in {0,...,4}{\foreach \c in {0,...,4}{
  \pgfmathsetmacro{\sh}{int(12+30*abs(sin(47*(\r+1)*(\c+2))))}
  \draw[black!40, fill=palOrange!\sh] ($(P3.north west)+(0.24+\c*0.2,-1.12-\r*0.2)$) rectangle +(0.2,0.2);}}
\foreach \r in {0,...,4}{\draw[black!40, fill=white] ($(P3.north west)+(1.50,-1.12-\r*0.2)$) rectangle +(0.2,0.2);
  \draw[black!40, fill=red!20] ($(P3.north west)+(2.16,-1.12-\r*0.2)$) rectangle +(0.2,0.2);}
\node[font=\tiny] at ($(P3.north west)+(1.93,-1.42)$) {$=$};
\node[note, anchor=north] at ($(P3.north west)+(0.74,-1.96)$) {interactions};
\node[note, anchor=north] at ($(P3.north west)+(1.60,-1.96)$) {$\vect{\lambda}$};
\node[note, anchor=north] at ($(P3.north west)+(2.26,-1.96)$) {deficits};
\draw[flow] ($(P3.north west)+(1.60,-2.30)$) -- ($(P3.south west)+(1.60,1.28)$);
\node[note, anchor=west] at ($(P3.north west)+(1.66,-2.62)$) {patch};
\foreach \c in {0,...,11}{\draw[cell] ($(P3.south west)+(0.21+\c*0.2,0.92)$) rectangle +(0.2,0.30);}
\foreach \c in {1,4,8,10}{\draw[sel] ($(P3.south west)+(0.21+\c*0.2,0.92)$) rectangle +(0.2,0.30);}
\pic[scale=1.15, rotate=-20] at ($(P3.north west)+(1.14,-2.64)$) {bug};
\pic[rotate=-30, scale=0.9] at ($(P3.north west)+(1.14,-2.64)$) {bandage};
\draw[sel] ($(P3.south west)+(0.16,0.42)$) rectangle +(0.2,0.22); \node[note, anchor=west] at ($(P3.south west)+(0.42,0.53)$) {patched write vector};
\draw[cell] ($(P3.south west)+(0.16,0.13)$) rectangle +(0.2,0.22); \node[note, anchor=west] at ($(P3.south west)+(0.42,0.24)$) {unchanged write vector};
\coordinate (O) at ($(P3.north east)+(0.50,0)$);
\node[hdr] at ($(O)+(0.10,-0.24)$) {Output};
\pic[scale=1.15] at ($(O)+(0.66,\mid)$) {lm};
\pic[rotate=-30, scale=0.9] at ($(O)+(0.66+0.391,\mid+0.345)$) {bandage};
\node[cnote, anchor=north] at ($(O)+(0.66,\mid-0.54)$) {repaired\\language model};
\draw[wide] (1.54,\mid) -- (P1.west |- 0,\mid);
\draw[wide] (P1.east |- 0,\mid) -- (P2.west |- 0,\mid);
\draw[wide] (P2.east |- 0,\mid) -- (P3.west |- 0,\mid);
\draw[wide] (P3.east |- 0,\mid) -- ++(0.50,0);
\node[tag=palBlue, anchor=south] at ($(P2.north)+(0,0.08)$) {targeting gap};
\node[tag=palOrange, anchor=south] at ($(P3.north)+(0,0.08)$) {tailoring gap};
\draw[wide, rounded corners=4pt] (P3.south) -- ++(0,-0.30) -| (P1.south);
\node[inner sep=0pt, anchor=west, font=\scriptsize] at ($(P1.south west)+(0.12,-0.62)$) {\textbf{Re-linearize the Repaired Model}\quad{\tiny\color{black!75}repeat locating and tailoring until no token fails}};
\end{tikzpicture}
\caption{Overview of \astra{} on a running example. The three panels are the steps of one round, and the loop below them repeats the first and the third step.}
\label{fig:overview}
\end{figure}
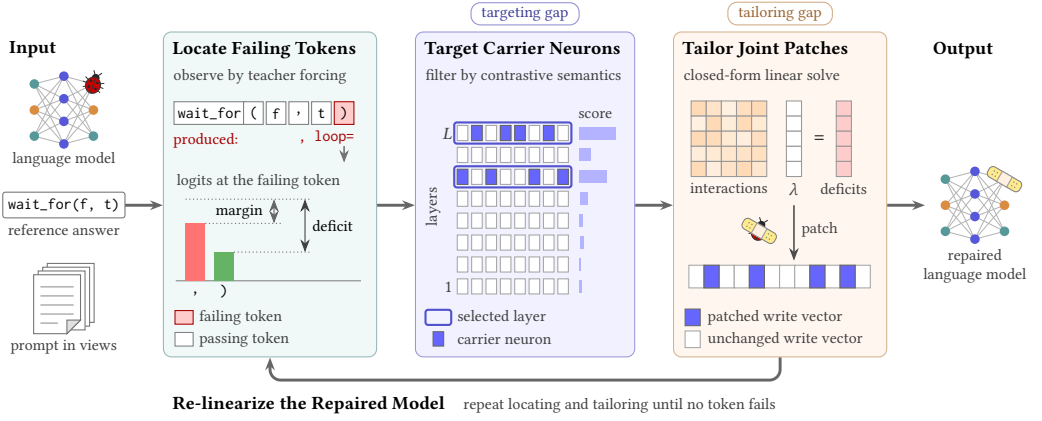

This section presents \astra{}, which repairs one failing sample at a time. Figure~\ref{fig:overview} gives an overview with a running example, in which a release has removed the parameter \texttt{loop} of \texttt{wait\_for} and the model still passes it. The input is a repair prompt, rendered in four ways that we call views, and a reference answer. \astra{} first observes the model on every view and locates the answer tokens at which it fails, here the closing parenthesis, to which the model prefers a comma (Section~\ref{sec:observe}). It then targets the carrier neurons, selected by \semcon{} in late layers (Section~\ref{sec:attribute}), which addresses the targeting gap. Next, it tailors a joint patch for all failing tokens by solving one linear system in closed form and writes it into the write vectors of the carriers (Section~\ref{sec:solve}), which addresses the tailoring gap. Finally, it re-linearizes the repaired model by repeating the first and the third step (Section~\ref{sec:relinearize}). The output is the repaired model, which differs from the original only in the write vectors of the carrier neurons.

\subsection{Preliminaries}
\label{sec:prelim}

Consider a decoder-only language model with $L$ layers. It keeps a residual stream at every position, to which each layer adds the output of its attention block and of its feed-forward block. The feed-forward block of layer $l$ computes one scalar per intermediate unit, the key activation $k^{l}_{j}$ of neuron $(l,j)$. In the gated blocks of the three models, it is the product of the activated gate projection and the up projection of the block input. The output projection $W^{l}_{\text{down}}\in\mathbb{R}^{H\times N}$ then adds $\sum_j k^{l}_{j}\,\vect{w}^{l}_j$ to the residual stream, where the write vector $\vect{w}^{l}_j$ is column $j$ of $W^{l}_{\text{down}}$. After the last layer, the final normalization divides the residual stream by its root mean square, the scale, and multiplies it by a gain $\vect{g}$, and the output embedding $W\in\mathbb{R}^{V\times H}$ maps the result to the logits of the $V$ tokens.

A \emph{carrier neuron}, or carrier for short, is a feed-forward neuron whose write vector, a column of the output projection, is changed by the patch, and it carries the patch into the residual stream whenever it is active. The carriers form the set $\mathcal{S}$, of which $\mathcal{S}_l$ lie in layer $l$. The write vector is the unit of repair for two reasons. First, a changed write vector alters only what its neuron writes when it is active. Second, the logits are linear in the final residual stream up to the final normalization, so the first-order effect of the change on the lead of one token over another is known exactly.

The repair is expressed in vocabulary-defined semantics~\cite{gu2024vds,gu2025star}. The output embedding assigns every token a direction of the hidden space, its row of $W$, so the logits are the coordinates of a hidden state in these vocabulary-defined semantic directions. Since every write vector lies in the same space, what a neuron writes can be read in the same coordinates~\cite{geva2022promoting}. A failure concerns two tokens, the \emph{target token} of the reference answer and the \emph{produced token} that the model prefers to it, and the contrast between their two directions is a semantic coordinate. \Semcon{} scores a neuron by what it currently contributes to this coordinate.

Finally, the repair sample consists of a repair prompt $q$ and a reference answer $\vect{y}=(y_t)_{t\le T}$ of $T$ tokens. A patch fitted to one phrasing is tied to the key activations of that phrasing, whereas the test prompt is worded differently. \astra{} therefore uses four views $q^{(1)},\dots,q^{(4)}$ of the repair prompt, namely the original prompt, a Markdown document with sections, a narrative and a single instruction sentence, all with the same content and the same reference answer. The indices $t$ and $u$ run over the answer tokens of all views, so an answer token counts once per view, and a quantity with the index $t$ is taken at the position that predicts this token.

\subsection{Locating Failing Tokens}
\label{sec:observe}

The first step locates the tokens at which the model fails on the repair sample and determines what would correct each failure at the output. \astra{} observes the model under teacher forcing, that is, in one forward pass per view over the view followed by the reference answer. The prediction of every answer token $y_t$ is then conditioned on the correct prefix, so each token is judged independently of earlier errors. For every target token $y_t$, the observation records its strongest competitor $p_t$, the token with the highest logit other than $y_t$, which is the produced token whenever the model errs.

A \emph{failing token} is an answer token whose logit lead over its competitor is below a margin $\mu$, that is, a wrongly predicted token or a correct one with a narrow lead. Its deficit $\gamma_t$ is the margin minus the lead. The margin is required because a tie holds only under the conditions of the observation. The repaired model later generates freely from a differently worded test prompt, where a small shift of the residual stream would reverse a lead of zero. The set $\mathcal{F}$ of failing tokens spans all views, and each of them keeps its own key activations $k^{l}_{j,t}$, scale $s_t$ and competitor.

The quantity to change for a failing token is its lead, because this contrast between two tokens decides the prediction. Since the logits are linear in the normalized final state, the lead is a linear function of the final residual stream. The \emph{contrast covector} expresses this function as the difference of the two rows of the output embedding, scaled by the gain:
\[
\vect{c}_t=\bigl(W_{y_t}-W_{p_t}\bigr)\odot \vect{g},
\]
where $W_{y}$ is the row of the output embedding for token $y$. A vector $\vect{v}$ that is added to the final residual stream at position $t$ thus changes the lead by $\vect{c}_t\!\cdot\!\vect{v}/s_t$.

The contrast covector states how a written vector is read, not which vector should be written. Many vectors have the same product with $\vect{c}_t$, and they differ in how much they move the logits of all other tokens. For every failing token, \astra{} therefore chooses the \emph{preferred direction} $\vect{d}_t$, the vector that raises the contrast by one unit with the least change of the full logit vector. With the output metric $G=W_g^{\top}W_g$ of the scaled output embedding $W_g=W\,\mathrm{diag}(\vect{g})$, this direction is $\vect{d}_t=G^{-1}\vect{c}_t/(\vect{c}_t^{\top}G^{-1}\vect{c}_t)$, the minimizer of $\lVert W_g\vect{d}\rVert^2$ subject to $\vect{c}_t\!\cdot\!\vect{d}=1$. Up to its scale, it is the difference between the latent representations that the pseudo-inverse of $W_g$ assigns to the two tokens, which also provide the target of the neuron patches of STAR~\cite{gu2024vds,gu2025star}.

\subsection{Targeting Carrier Neurons}
\label{sec:attribute}

The second step targets the neurons that should carry the patch, namely those that can move the failing contrasts with a small change. The attribution follows the path on which a patch acts. At the position of token $t$, neuron $(l,j)$ adds $k^{l}_{j,t}\vect{w}^{l}_j$ to the residual stream, and the output reads this vector through the contrast covector. The direct contribution of the neuron to the lead is therefore the product of how active the neuron is and how far its write vector points along the contrast. \Semcon{} scores a neuron by its contribution to the logit contrast between the target token and the produced token in vocabulary-defined semantic coordinates. The score sums the size of this product over the failing tokens:
\[
\mathrm{CS}^{l}_{j}=\sum_{t\in\mathcal{F}}\Bigl\lvert \frac{k^{l}_{j,t}\,(\vect{c}_t\!\cdot\!\vect{w}^{l}_j)}{s_t}\Bigr\rvert .
\]

Three properties make \semcon{} suitable for selecting carriers. First, it is forward-only, because the key activations, the scales and the covectors come from the forward pass that finds the failing tokens. Second, it is additive, because the contributions of the neurons of a layer sum to the direct contribution of its feed-forward block, so one quantity ranks both neurons and layers. Third and most important, it is measured in the coordinates of the patch: a change of the write vector acts on the lead through the same key activation and the same covector, so a neuron with a high score needs only a small change. The sum takes absolute values because a neuron that writes against the target is as useful a carrier as one that writes toward it, since the patch can redirect both, and because signed contributions to different failing tokens would cancel. The measure is a contrast-specific form of direct logit attribution~\cite{elhage2021mathematical,nostalgebraist2020logitlens} and, unlike the attributions of Neuron Patching and STAR~\cite{gu2023neuronpatching,gu2025star}, needs neither a gradient nor the occlusion of layers.

The selection proceeds from layers to neurons. \astra{} ranks the layers by their summed scores and keeps the quarter of the layers with the highest sums. Within each kept layer, it keeps the neurons with the highest scores. The number of kept neurons grows with the answer, because every answer token can fail in every view and each failing token adds one requirement to the patch. The budget is $R=16$ neurons per answer token, spread evenly over the $L_{\mathcal{S}}$ kept layers, which gives $\lceil R\,T/L_{\mathcal{S}}\rceil$ neurons per layer. This number is bounded below by 16, so that short answers retain enough carriers, and above by 4 percent of the width $N$, so that the patch stays sparse in every layer. The selection is made once, in the first round, because the attribution shifts as soon as a patch is applied, and a new selection in every round would add further neurons, whereas a fixed set keeps the patch within its budget and lets later rounds refine the same columns.

\subsection{Tailoring Joint Patches}
\label{sec:solve}

The third step tailors the patch, that is, it computes what the carrier neurons should write. The requirement is that the lead of every failing token grows by its deficit. Through the direct path, a change $\delta\vect{w}^{l}_j$ of a write vector changes the lead at token $t$ by $k^{l}_{j,t}\,(\vect{c}_t\!\cdot\!\delta\vect{w}^{l}_j)/s_t$, so each requirement is linear in the patch. Since the requirements are far fewer than the entries of the selected columns, many patches satisfy them. Among these, \astra{} takes the one with the least weighted change, in which every carrier receives a combination of the preferred directions:
\[
\delta\vect{w}^{l}_j=D^{l}_j\sum_{u\in\mathcal{F}}\lambda_u\,k^{l}_{j,u}\,\vect{d}_u .
\]
The coefficient $\lambda_u$ is the strength of the correction for failing token $u$, and a carrier takes part in this correction in proportion to its key activation on that token. Each carrier thus receives its own patch, a mixture of the directions of the failing tokens on which it is active. The null-space weight $D^{l}_j$ protects unrelated behavior. It is a diagonal form of the null-space constraint of AlphaEdit~\cite{fang2025alphaedit}, namely the inverse of the mean squared key activation of the neuron on a set of reference prompts. Neurons that are active on unrelated inputs thus carry little of the patch.

\begin{algorithm}[htb]
\caption{\astra{} for one repair sample}
\label{alg:astra}
\small
\begin{algorithmic}[1]
\Require model $f_\theta$, views $q^{(1)},\dots,q^{(4)}$ of the repair prompt, reference answer $\vect y$, null-space weights $D$
\For{round $r=1,\dots,10$}
  \State observe $f_\theta$ on every view with teacher forcing, and collect the failing tokens $\mathcal F$ with their key activations $k^l_{j,t}$, scales $s_t$ and deficits $\gamma_t$
  \If{$\mathcal F=\emptyset$} \textbf{break} \EndIf
  \State $\vect c_t\gets(W_{y_t}-W_{p_t})\odot\vect g$ and $\vect d_t\gets G^{-1}\vect c_t/(\vect c_t^{\top}G^{-1}\vect c_t)$ for every $t\in\mathcal F$
  \If{$r=1$} select the layers and the neurons $\mathcal S$ with the highest \semcon{} scores \EndIf
  \State solve $A\vect\lambda=\vect\gamma$ with $A_{tu}=(\vect c_t\!\cdot\!\vect d_u)\,\tfrac{1}{s_t}\sum_{l}\sum_{j\in\mathcal S_l}D^l_j k^l_{j,t}k^l_{j,u}$
  \State $\vect w^l_j\gets\vect w^l_j+D^l_j\sum_{u\in\mathcal F}\lambda_u k^l_{j,u}\vect d_u$ for every $(l,j)\in\mathcal S$
\EndFor
\State \Return the repaired model
\end{algorithmic}
\end{algorithm}

The coefficients follow from the requirements. Inserting the patch into the change of every lead gives one linear equation per failing token, and together they form the system $A\vect{\lambda}=\vect{\gamma}$ with
\[
A_{tu}=(\vect{c}_t\!\cdot\!\vect{d}_u)\;\frac{1}{s_t}\sum_{l}\sum_{j\in\mathcal{S}_l}D^{l}_j\,k^{l}_{j,t}\,k^{l}_{j,u}.
\]
The entry $A_{tu}$ is the effect of the correction for token $u$ on the lead of token $t$. It is large when the two tokens activate the same carrier neurons and the direction of one is read by the covector of the other. These entries are the reason for a joint solve: an update computed for one failing token moves the leads of all tokens that share its carriers, so a token-by-token procedure undoes earlier corrections, whereas the joint system accounts for all interactions at once.

The system is small, with one row per failing token, and it is solved directly. The patch is analytic in this sense: it needs no optimizer, no step size and no backward pass. A small ridge term, a fraction of the mean diagonal entry, is added to the diagonal, because nearly identical requirements, such as the same token failing in two views, make $A$ ill-conditioned. The patch changes only the selected columns of $W^{l}_{\text{down}}$ in the kept layers.

\subsection{Re-linearizing the Repaired Model}
\label{sec:relinearize}

The fourth step re-linearizes the repaired model, because the solution of the linear system is exact only at first order. The system describes the direct path from the carrier neurons to the output. In the model, however, the later layers react to the changed residual stream, the scale of the final normalization moves, and the strongest competitor of a token may change. After each solve, the patch is therefore written into the weights and all views are observed again. The tokens that still fail form the next system, with new deficits, covectors and key activations over the same carrier neurons. The loop stops as soon as no token fails in any view, or after ten rounds, and the final patch sums the solutions of all rounds. Algorithm~\ref{alg:astra} lists the procedure, in which lines 2 to 4 are the observation, line 5 is the selection of the first round, and lines 6 and 7 are the solve and the patch.

\section{Experiments}
\label{sec:setup}

The experiments answer four research questions.
\begin{itemize}
\item \RQ{1}{Targeting and Tailoring Gaps} How far do the neurons that failure attribution targets differ from the carriers of a patch, and how far is the patch that they need from a generic update?
\item \RQ{2}{Selection of Carrier Neurons} How should the carrier neurons be selected to address the targeting gap, and how strongly and in which layers does the patch then change them?
\item \RQ{3}{Effectiveness and Efficiency} How effective and how fast is \astra{} against model editing, semantic analysis and low-rank adaptation, and which of its components matter most?
\item \RQ{4}{Side Effects and Robustness} How much does a repair change unrelated code, and how robust is the advantage of \astra{} to an unseen phrasing of the test prompt?
\end{itemize}

\paragraph{Baselines.} We compare with two state-of-the-art methods that have not been compared directly, one from each family, and with low-rank adaptation as a training reference. \emph{AlphaEdit}~\cite{fang2025alphaedit} represents model editing. It extends the locate-then-edit line of ROME and MEMIT with a null-space constraint, is applied to layers 4 to 8, and estimates its covariance from 1000 WikiText samples. \emph{STAR}~\cite{gu2025star} represents semantic analysis. It is the successor of Neuron Patching~\cite{gu2023neuronpatching}, selects a quarter of the layers by occlusion of their feed-forward blocks, and updates them in batch mode for eight epochs. \emph{LowRank} adapts the feed-forward output projections with low-rank adapters~\cite{hu2022lora} of rank 8, trained with AdamW for ten steps on the repair prompt and the reference answer. \emph{Unrepaired} is the model without a repair. All baselines use their default hyperparameters, and all methods are given the same repair pair. Extensions of editing to long-form or graph-structured knowledge~\cite{jiang2025anyedit,zhang2024glame} are orthogonal to carrier selection, and retrieval leaves the model unchanged, so we exclude both.

\paragraph{Benchmarks.} We use two executable benchmarks, PyEvo with 622 Python tasks and RustEvo with 708 Rust tasks. Each task pairs an API change (a deprecation, a signature change, a stabilized function or an implicit behavior change) with a programming task that needs the changed interface, a reference solution and an executable test program. RustEvo is derived from the Rust API evolution benchmark of prior work~\cite{liang2025rustevo}, and PyEvo applies the same construction to changes of the Python standard library. Every task has a \emph{repair pair} and a \emph{test prompt}. The repair pair consists of a repair prompt, which states the new API information and a rephrased task, and the reference solution as its reference answer. The test prompt poses the original task without any API information. A method thus sees only the repair pair, and the repaired model answers the test prompt.

\paragraph{Models.} We repair three open code models. The \emph{small model} is MiniCPM5-1B~\cite{hu2024minicpm}, which has \layersMo{} layers. The two \emph{large models} are OpenCoder-8B-Instruct~\cite{huang2024opencoder}, called OpenCoder-8B below, and OxCoder-9B, a hybrid-architecture model from the Hugging Face hub. Both have \layersOp{} layers and differ in architecture and training data. A \emph{setting} is one model on one benchmark, which gives \cellsAllTotal{} settings. All models run in bfloat16 on one GPU with greedy decoding.

\paragraph{Protocol.} Each sample is repaired independently from the original weights, and the answer of the repaired model is judged by running the tests. Pass@1 is the share of samples whose answer passes all tests. The main comparison of \rqtext{3} uses the complete benchmarks in all \cellsAllTotal{} settings, whereas the gap statistics of \rqtext{1}, the comparison of attributions of \rqtext{2}, the ablation of \rqtext{3} and the robustness study of \rqtext{4} use a \emph{shared subset}, the same samples for all runs. Each benchmark is divided into twelve interleaved parts that preserve the shares of the types of API change. The shared subset consists of four parts, one third of each benchmark (208 PyEvo and 236 RustEvo samples). Differences between two methods are tested with an exact McNemar test on the paired outcomes of the same samples, and intervals are 95 percent bootstrap intervals. For \rqtext{3}, the repair time is the wall-clock time per sample without the generation of the answer, and for \rqtext{4}, side effects follow the protocol of STAR on HumanEval~\cite{chen2021codex}: each of \ovRepairs{} problems is repaired in turn with its canonical solution as the reference answer, and after every repair the \ovHeldOut{} held-out problems are scored by teacher forcing against the unrepaired model.

\paragraph{Configuration.} \astra{} keeps the quarter of the layers with the highest summed \semcon{} scores and uses a margin of $\mu=5$ logits, \numViews{} views, a budget of $R=16$ neurons per answer token and at most ten rounds. It thus keeps \keptMo{}, \keptOp{} and \keptOx{} layers of the three models. The null-space weights are estimated on 32 test prompts of the other benchmark, so no evaluated sample is among them. A repair changes between \neuronShareSmMin{} and \neuronShareSmMax{} percent of all feed-forward neurons on the small model and between \neuronShareMin{} and \neuronShareMax{} percent on the large ones, on average over the samples of a setting. This configuration was chosen on a development split of every eighth sample of each benchmark with the large models and then kept unchanged, including for the small model. Without the development samples, \astra{} still reaches \testAllAvgAstra{} percent Pass@1 on average over all \cellsAllTotal{} settings against \testAllBest{} percent for the best baseline (\testAllBestName{}).

\subsection{Targeting and Tailoring Gaps}
\label{sec:gap}

To measure the two gaps of \rqtext{1}, we compute for each sample the unconstrained closed-form patch that corrects all failing tokens, in which every neuron of a layer may change. Its \emph{natural carriers} are the neurons whose entries contribute most to the patch. This set is a reference derived from the solver, not a ground truth. We compare the natural carriers with the ranking by failure attribution in four statistics, computed in the layers that \astra{} selects. The \emph{rank correlation} is the Spearman correlation between the attribution score and the contribution of a neuron to the patch, and the \emph{top-set overlap} is the Jaccard index of the two top sets whose size equals the neuron budget of the repair. These two statistics measure the targeting gap. The \emph{patch cosine} is the mean cosine between the patches that the unconstrained solution assigns to the 32 most useful carriers of a layer. It measures the tailoring gap, since it shows whether one shared vector could serve them. Finally, the \emph{shift} is one minus the top-set overlap of the attribution before and after the first round of the patch, which shows how the two gaps are coupled. The failure attribution in these statistics is \semcon{}, which is closest to the patch, so the values are a favorable case.

\begin{table}[htb]
\centering\tabsmall
\caption{Gap statistics between failure attribution and the natural carriers of the patch, averaged over the shared subset of each setting. Tokens counts the failing tokens per sample, Correlation the rank correlation, Overlap the top-set overlap, Cosine the patch cosine, and Shift the attribution shift after the first round. Correlation and Overlap quantify the targeting gap, Cosine the tailoring gap, and Shift the coupling of the two gaps. The arrows point toward a smaller gap.}
\label{tab:gap}
\begin{tabular}{llrrrrr}
\toprule
\textbf{Model} & \textbf{Benchmark} & \textbf{Tokens} & \textbf{Correlation $\uparrow$} & \textbf{Overlap $\uparrow$} & \textbf{Cosine $\uparrow$} & \textbf{Shift $\downarrow$} \\
\midrule
MiniCPM5-1B & PyEvo & 293 & 0.50 & 0.17 & +0.001 & 0.11 \\
 & RustEvo & 122 & 0.53 & 0.18 & +0.009 & 0.14 \\
\midrule
OpenCoder-8B & PyEvo & 247 & 0.50 & 0.18 & $-$0.001 & 0.13 \\
 & RustEvo & 92 & 0.49 & 0.15 & $-$0.001 & 0.17 \\
\midrule
OxCoder-9B & PyEvo & 221 & 0.37 & 0.20 & 0.000 & 0.10 \\
 & RustEvo & 85 & 0.39 & 0.19 & +0.002 & 0.12 \\
\bottomrule
\end{tabular}

\end{table}

Table~\ref{tab:gap} shows the targeting gap in every setting: failure attribution is informative about the carriers of a patch but does not identify them. The rank correlation with the natural carriers is moderate, between \rqCorrMin{} and \rqCorrMax{}, whereas the top-set overlap is much lower, between \rqOverMin{} and \rqOverMax{}. The highest-ranked neurons and the natural carriers thus share at most \rqOverPctMax{} percent of their union on average. As a top set holds at most 4 percent of the neurons of a layer, two unrelated sets of this size would share about 2 percent of their union, so the attribution is far from random and far from sufficient. The \semcon{} score of a neuron is its key activation times the alignment of its current write vector with the contrast, whereas the share of a neuron in the patch depends on its key activations and its null-space weight, not on where its write vector points. The shared key activation yields the moderate correlation, but the neurons at the top of the attribution already write strongly along the contrast, which no patch requires.

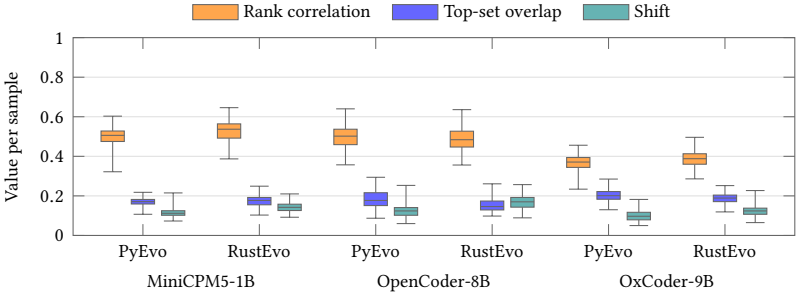
\begin{figure}[htb]
\centering
\begin{tikzpicture}
\begin{axis}[paperfig, boxplot/draw direction=y, ymin=0, ymax=1,
  ytick={0,0.2,0.4,0.6,0.8,1}, ylabel={Value per sample},
  xtick={1,2,3,4,5,6}, xticklabels={PyEvo,RustEvo,PyEvo,RustEvo,PyEvo,RustEvo}, xmin=0.4, xmax=6.6, after end axis/.code={\node[font=\scriptsize, anchor=north] at (rel axis cs:0.1774,-0.15) {MiniCPM5-1B};\node[font=\scriptsize, anchor=north] at (rel axis cs:0.5000,-0.15) {OpenCoder-8B};\node[font=\scriptsize, anchor=north] at (rel axis cs:0.8226,-0.15) {OxCoder-9B};}]
\addlegendimage{area legend, fill=sCorr, draw=black!60}\addlegendentry{Rank correlation}
\addlegendimage{area legend, fill=sOver, draw=black!60}\addlegendentry{Top-set overlap}
\addlegendimage{area legend, fill=sShift, draw=black!60}\addlegendentry{Shift}
\addplot[forget plot, fill=sCorr, draw=black!60, line width=0.4pt, boxplot prepared={draw position=0.74, box extend=0.2, lower whisker=0.322, lower quartile=0.475, median=0.506, upper quartile=0.528, upper whisker=0.603}] coordinates {};
\addplot[forget plot, fill=sOver, draw=black!60, line width=0.4pt, boxplot prepared={draw position=1.00, box extend=0.2, lower whisker=0.107, lower quartile=0.159, median=0.171, upper quartile=0.182, upper whisker=0.218}] coordinates {};
\addplot[forget plot, fill=sShift, draw=black!60, line width=0.4pt, boxplot prepared={draw position=1.26, box extend=0.2, lower whisker=0.073, lower quartile=0.102, median=0.112, upper quartile=0.125, upper whisker=0.215}] coordinates {};
\addplot[forget plot, fill=sCorr, draw=black!60, line width=0.4pt, boxplot prepared={draw position=1.74, box extend=0.2, lower whisker=0.387, lower quartile=0.492, median=0.537, upper quartile=0.564, upper whisker=0.646}] coordinates {};
\addplot[forget plot, fill=sOver, draw=black!60, line width=0.4pt, boxplot prepared={draw position=2.00, box extend=0.2, lower whisker=0.103, lower quartile=0.155, median=0.177, upper quartile=0.192, upper whisker=0.249}] coordinates {};
\addplot[forget plot, fill=sShift, draw=black!60, line width=0.4pt, boxplot prepared={draw position=2.26, box extend=0.2, lower whisker=0.092, lower quartile=0.126, median=0.142, upper quartile=0.158, upper whisker=0.210}] coordinates {};
\addplot[forget plot, fill=sCorr, draw=black!60, line width=0.4pt, boxplot prepared={draw position=2.74, box extend=0.2, lower whisker=0.357, lower quartile=0.459, median=0.502, upper quartile=0.537, upper whisker=0.640}] coordinates {};
\addplot[forget plot, fill=sOver, draw=black!60, line width=0.4pt, boxplot prepared={draw position=3.00, box extend=0.2, lower whisker=0.087, lower quartile=0.151, median=0.177, upper quartile=0.216, upper whisker=0.294}] coordinates {};
\addplot[forget plot, fill=sShift, draw=black!60, line width=0.4pt, boxplot prepared={draw position=3.26, box extend=0.2, lower whisker=0.060, lower quartile=0.102, median=0.124, upper quartile=0.141, upper whisker=0.253}] coordinates {};
\addplot[forget plot, fill=sCorr, draw=black!60, line width=0.4pt, boxplot prepared={draw position=3.74, box extend=0.2, lower whisker=0.356, lower quartile=0.447, median=0.484, upper quartile=0.527, upper whisker=0.636}] coordinates {};
\addplot[forget plot, fill=sOver, draw=black!60, line width=0.4pt, boxplot prepared={draw position=4.00, box extend=0.2, lower whisker=0.098, lower quartile=0.129, median=0.146, upper quartile=0.174, upper whisker=0.261}] coordinates {};
\addplot[forget plot, fill=sShift, draw=black!60, line width=0.4pt, boxplot prepared={draw position=4.26, box extend=0.2, lower whisker=0.089, lower quartile=0.143, median=0.170, upper quartile=0.192, upper whisker=0.257}] coordinates {};
\addplot[forget plot, fill=sCorr, draw=black!60, line width=0.4pt, boxplot prepared={draw position=4.74, box extend=0.2, lower whisker=0.234, lower quartile=0.344, median=0.371, upper quartile=0.394, upper whisker=0.456}] coordinates {};
\addplot[forget plot, fill=sOver, draw=black!60, line width=0.4pt, boxplot prepared={draw position=5.00, box extend=0.2, lower whisker=0.130, lower quartile=0.183, median=0.201, upper quartile=0.222, upper whisker=0.285}] coordinates {};
\addplot[forget plot, fill=sShift, draw=black!60, line width=0.4pt, boxplot prepared={draw position=5.26, box extend=0.2, lower whisker=0.050, lower quartile=0.079, median=0.097, upper quartile=0.117, upper whisker=0.182}] coordinates {};
\addplot[forget plot, fill=sCorr, draw=black!60, line width=0.4pt, boxplot prepared={draw position=5.74, box extend=0.2, lower whisker=0.286, lower quartile=0.360, median=0.388, upper quartile=0.413, upper whisker=0.496}] coordinates {};
\addplot[forget plot, fill=sOver, draw=black!60, line width=0.4pt, boxplot prepared={draw position=6.00, box extend=0.2, lower whisker=0.119, lower quartile=0.171, median=0.189, upper quartile=0.204, upper whisker=0.252}] coordinates {};
\addplot[forget plot, fill=sShift, draw=black!60, line width=0.4pt, boxplot prepared={draw position=6.26, box extend=0.2, lower whisker=0.065, lower quartile=0.107, median=0.124, upper quartile=0.138, upper whisker=0.227}] coordinates {};
\end{axis}
\end{tikzpicture}
\caption{Distribution of the rank correlation, the top-set overlap and the shift over the samples of the shared subset of each setting. Boxes show the median and the quartiles, and whiskers the extremes.}
\label{fig:gaps}
\end{figure}

Since averages could hide samples in which attribution and patch agree, Figure~\ref{fig:gaps} shows the distribution of the statistics over the samples. The targeting gap holds for every single sample: among the \rqSamples{} samples of the \rqSettings{} settings, the top-set overlap never exceeds \rqOverSampleMax{}, and its interquartile range spans at most \rqOverIqrMax{} in a setting. The rank correlation, in turn, lies between \rqCorrSampleMin{} and \rqCorrSampleMax{}, so the attribution is informative for every sample and sufficient for none. Moreover, OxCoder-9B has the lowest rank correlation, with medians of \rqCorrMedOxMin{} and \rqCorrMedOxMax{}, and yet the highest top-set overlap, with medians of \rqOverMedOxMin{} and \rqOverMedOxMax{}. Agreement over the whole ranking thus does not translate into agreement at its top, where the carriers are chosen.

The tailoring gap appears in the patch cosine, which is at most \rqCosAbs{} in absolute value on average in every setting and at most \rqCosSampleAbs{} in any sample, so the patches of different carriers are nearly orthogonal. Each carrier receives a combination of the directions of the tokens on which it is active, and with \rqTokMin{} to \rqTokMax{} failing tokens per sample on average, the combinations differ between carriers. Hence, no rule that moves all selected neurons along one direction can represent such patches.

The shift shows that the two gaps are coupled, since the attribution changes when the patch is applied. After one round, the shift is \rqShiftMin{} to \rqShiftMax{} on average and between \rqShiftSampleMin{} and \rqShiftSampleMax{} in single samples, so the top set of the attribution overlaps with its former self by only \rqKeepMin{} to \rqKeepMax{} on average. The patch alters the write vectors that the attribution reads and, through them, the activations of the later layers. A ranking that is computed once therefore describes the unrepaired model and not the model that receives the patch, which is why \astra{} fixes its carriers in the first round and then solves for all failing tokens jointly. The loss of a repair that instead corrects the failing tokens one by one is measured in the ablation study.

The statistics are similar for the small model and the large ones. However, the targeting gap does not imply that a selection by attribution must fail, since the requirements of a patch are far fewer than the entries of the candidate columns and many sets of carriers can satisfy them. Rather, it implies that the neurons at the top of an attribution are not by themselves suitable carriers, so an attribution has to be judged by the patch that its carriers require and by the repair that follows.

\answer{1}{The neurons that failure attribution targets and the carriers of a patch agree moderately in rank (correlation \rqCorrMin{} to \rqCorrMax{}), but their top sets overlap by only \rqOverMin{} to \rqOverMax{} on average and by at most \rqOverSampleMax{} in any sample (targeting gap). The patches that the carriers need are nearly orthogonal (cosine at most \rqCosAbs{}) and thus far from a generic update (tailoring gap). The gaps are coupled, since one round of patching shifts the targeted top set by \rqShiftMin{} to \rqShiftMax{}.}

\subsection{Selection of Carrier Neurons}
\label{sec:rq2}

For \rqtext{2}, addressing the targeting gap means selecting neurons that can carry the patch, that is, neurons that repair the failure under a small change of their write vectors, while the top-set overlap itself stays unchanged. To separate the effect of the attribution from that of the solver, all variants use the same closed-form patch and the same neuron budget and differ only in the score that ranks the neurons. We compare \semcon{} with two attributions that use a backward pass. \ixg{} scores a neuron by its key activation times the gradient of the cross-entropy of the answer. Contrastive gradient, in turn, applies the same product to the logit contrast of the failing tokens~\cite{yin2022contrastive} and thus accounts for the effect of the neuron through all later layers. It scores the same contrast as \semcon{}, but through a gradient and not through the semantic coordinates of the output. For every repair, we also log two quantities over the neurons that the patch changes. The \emph{relative patch norm} is the norm of the change of a write vector divided by the norm of the write vector, and the \emph{parallelism} is the absolute cosine between the change and the write vector.

\begin{table}[htb]
\centering\tabsmall
\caption{Pass@1 in percent and median relative patch norm of three attributions under the same closed-form patch and neuron budget, on the shared subset. Input$\times$Grad., Cont. grad. and Cont. sem. denote \ixg{}, contrastive gradient and \semcon{}. The best Pass@1 of each setting is in bold.}
\label{tab:attrib}
\begin{tabular}{llrrrrrr}
\toprule
 &  & \multicolumn{3}{c}{\textbf{Pass@1 $\uparrow$}} & \multicolumn{3}{c}{\textbf{Relative patch norm $\downarrow$}} \\
\cmidrule(lr){3-5}\cmidrule(lr){6-8}
\textbf{Model} & \textbf{Benchmark} & \textbf{Input$\times$Grad.} & \textbf{Cont. grad.} & \textbf{Cont. sem.} & \textbf{Input$\times$Grad.} & \textbf{Cont. grad.} & \textbf{Cont. sem.} \\
\midrule
MiniCPM5-1B & PyEvo & 29.8 & 35.6 & \textbf{37.0} & 0.19 & 0.16 & 0.10 \\
 & RustEvo & 64.0 & 68.6 & \textbf{74.6} & 0.10 & 0.09 & 0.07 \\
\midrule
OpenCoder-8B & PyEvo & 25.0 & 27.9 & \textbf{61.1} & 0.55 & 0.46 & 0.17 \\
 & RustEvo & 78.8 & 78.0 & \textbf{89.0} & 0.37 & 0.35 & 0.24 \\
\midrule
OxCoder-9B & PyEvo & 40.4 & 42.3 & \textbf{46.2} & 0.30 & 0.26 & 0.09 \\
 & RustEvo & 82.6 & \textbf{85.6} & 80.5 & 0.23 & 0.19 & 0.08 \\
\bottomrule
\end{tabular}

\end{table}

Table~\ref{tab:attrib} shows that the benefit of \semcon{} depends on the model. On OpenCoder-8B, it exceeds both gradient-based attributions, with \attrOpDir{} against \attrOpIxg{} and \attrOpPath{} percent over the two benchmarks, and the difference is significant in both settings of this model (paired McNemar test, $p<0.05$). On MiniCPM5-1B, it is the best attribution in both settings, significantly above both gradient-based attributions on RustEvo (\attrMoRuDir{} against \attrMoRuIxg{} and \attrMoRuPath{} percent) and above \ixg{} on PyEvo. On OxCoder-9B, however, the three attributions are close, with no significant difference in either setting, and both gradient-based attributions are higher on RustEvo (\attrOxRuIxg{} and \attrOxRuPath{} against \attrOxRuDir{} percent). Overall, it gives the highest Pass@1 in \attrCellsBest{} of the \attrCellsTotal{} settings, significantly above both alternatives in \attrCellsSig{} of them, and, pooled over all settings, lies \ablAllDIxg{} and \ablAllDPath{} points above \ixg{} and contrastive gradient. A backward pass through the later layers thus does not pay off, since the forward-only score is never significantly worse.

\begin{figure}[htb]
\centering
\begin{tikzpicture}
\begin{axis}[paperfig, paperbars, bar width=4.6pt, ymin=0, ymax=1,
  ytick={0,0.2,0.4,0.6,0.8,1}, ylabel={Relative patch norm},
  xtick={1,2,3,4,5,6}, xticklabels={PyEvo,RustEvo,PyEvo,RustEvo,PyEvo,RustEvo}, xmin=0.4, xmax=6.6, after end axis/.code={\node[font=\scriptsize, anchor=north] at (rel axis cs:0.1774,-0.15) {MiniCPM5-1B};\node[font=\scriptsize, anchor=north] at (rel axis cs:0.5000,-0.15) {OpenCoder-8B};\node[font=\scriptsize, anchor=north] at (rel axis cs:0.8226,-0.15) {OxCoder-9B};},
  error bars/y dir=both, error bars/y explicit, error bars/error bar style={line width=0.4pt, black!70}, error bars/error mark options={rotate=90, mark size=1.2pt, black!70}]
\addplot[fill=aIxg, draw=black!60] coordinates {(1,0.187) +- (0,0.332) -= (0,0.060) (2,0.097) +- (0,0.033) -= (0,0.018) (3,0.546) +- (0,0.137) -= (0,0.115) (4,0.369) +- (0,0.138) -= (0,0.087) (5,0.303) +- (0,0.121) -= (0,0.072) (6,0.232) +- (0,0.062) -= (0,0.044)};
\addlegendentry{Input$\times$Gradient}
\addplot[fill=aPath, draw=black!60] coordinates {(1,0.160) +- (0,0.125) -= (0,0.052) (2,0.091) +- (0,0.029) -= (0,0.016) (3,0.461) +- (0,0.125) -= (0,0.079) (4,0.351) +- (0,0.082) -= (0,0.062) (5,0.262) +- (0,0.066) -= (0,0.052) (6,0.193) +- (0,0.055) -= (0,0.044)};
\addlegendentry{Contrastive gradient}
\addplot[fill=aContrib, draw=black!60] coordinates {(1,0.096) +- (0,0.050) -= (0,0.025) (2,0.073) +- (0,0.016) -= (0,0.014) (3,0.168) +- (0,0.033) -= (0,0.024) (4,0.237) +- (0,0.033) -= (0,0.029) (5,0.086) +- (0,0.013) -= (0,0.012) (6,0.076) +- (0,0.013) -= (0,0.011)};
\addlegendentry{Contrastive semantics}
\end{axis}
\end{tikzpicture}
\caption{Relative patch norm under the three attributions on the shared subset of all \cellsAllTotal{} settings. Bars show the median over the repairs and whiskers the interquartile range. A lower norm means that the targeted neurons carry the patch under a smaller change, which addresses the targeting gap.}
\label{fig:geom}
\end{figure}
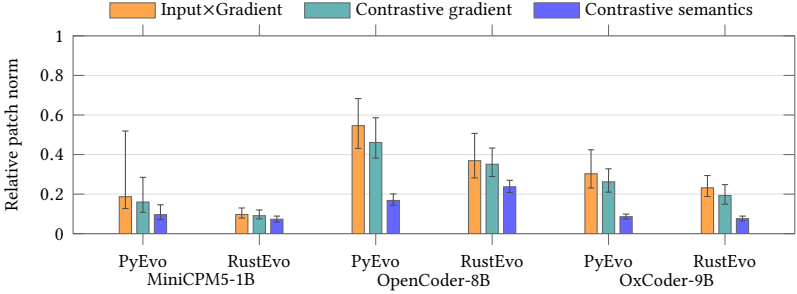

The size of the patch explains part of this dependence. As Figure~\ref{fig:geom} shows, selection by \semcon{} yields neurons that a patch changes less in all \relCellsLower{} settings. Under this selection, the median relative patch norm lies between \relDirMin{} and \relDirMax{}, whereas the gradient-based attributions reach between \relOthMin{} and \relOthMax{}. The neurons that \semcon{} targets can thus carry the patch at a small cost, which is what addressing the targeting gap requires. Per repair, the median ratio of the relative norm under a gradient-based attribution to that under \semcon{} lies between \ratioMin{} and \ratioMax{} over the settings. The largest patches coincide with the largest losses. In the two settings of OpenCoder-8B, where the gradient-based attributions lose the most, their median patch amounts to \relOpRuPath{} to \relOpPyIxg{} times the norm of the write vector, the highest values of all settings. A plausible reading is that a patch is exact only at first order: the larger the change of a write vector, the less the linear system describes the reaction of the later layers, and the rounds of re-linearization do not recover the difference. The relation is not monotone, however. On OxCoder-9B, patches of \relOxRuPath{} to \relOxPyIxg{} times the norm cause no significant loss, whereas on MiniCPM5-1B with RustEvo a significant loss occurs at about \relMoRuIxg{}. The relative patch norm thus indicates where the attribution matters most but does not predict a single repair.

The reason for the smaller patches lies in the coordinates of the scores. A gradient-based score ranks a neuron by its effect on the objective through all later layers, whereas the patch acts on the direct path from the neuron to the output. Only \semcon{} measures how far a neuron moves the contrast on this path. The parallelism shows in addition what kind of change a patch is. It is at most \parMax{} in any setting, so the patches turn the write vectors of the carriers and do not rescale them, and this holds for all attributions. A repair thus does not amplify or attenuate what a carrier already writes. It adds a direction that the carrier did not write before, which agrees with the low top-set overlap: neurons that already write along the contrast are not natural carriers.

\begin{figure}[htb]
\centering
\begin{tikzpicture}
\begin{axis}[paperfig, ymin=0, ymax=100, ytick={0,25,50,75,100}, xmin=0.60, xmax=1.01,
  xtick={0.6,0.7,0.8,0.9,1}, ylabel={Share of the repairs (percent)}, xlabel={Relative depth of the layer}]
\addplot[paperline, gMo, mark=triangle*] coordinates {(0.6250,0.2) (0.6667,0.0) (0.7083,1.0) (0.7500,80.2) (0.7917,99.4) (0.8333,54.1) (0.8750,65.2) (0.9167,100.0) (0.9583,100.0) (1.0000,100.0)};
\addplot[paperline, gOp, mark=*] coordinates {(0.6875,0.4) (0.7188,10.3) (0.7500,55.3) (0.7812,95.3) (0.8125,66.8) (0.8438,94.4) (0.8750,91.7) (0.9062,89.2) (0.9375,97.0) (0.9688,99.5) (1.0000,100.0)};
\addplot[paperline, gOx, mark=square*] coordinates {(0.7188,36.7) (0.7500,1.4) (0.7812,81.9) (0.8125,85.7) (0.8438,99.8) (0.8750,94.4) (0.9062,100.0) (0.9375,100.0) (0.9688,100.0) (1.0000,100.0)};
\legend{MiniCPM5-1B, OpenCoder-8B, OxCoder-9B}
\end{axis}
\end{tikzpicture}
\caption{Layers that \astra{} selects in the main comparison (\layerRepairsMo{} repairs per model). Each line shows the share of the repairs of a model that select a layer against its relative depth, the layer number divided by the number of layers. No repair selects a layer left of the plotted range.}
\label{fig:layers}
\end{figure}
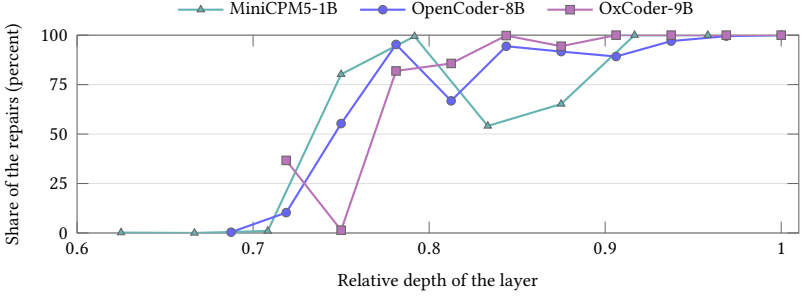

Figure~\ref{fig:layers} shows where in the depth \semcon{} places a patch. The selection is concentrated in the late layers of all three models. No layer in the first \layerNeverPct{} percent of the depth is selected in any model, and the last three layers are selected in at least \layerLastThreeMin{} percent of the repairs of every model. This outcome follows from the measure, which scores the direct contribution to the output, and from the role of late feed-forward layers, which write the vectors that the output embedding reads. It differs from locate-then-edit methods, which write into a fixed window of early layers. In between, the choice depends on the model and the failure. OpenCoder-8B uses layer 24 in \layerOpTwentyFour{} percent of the repairs, whereas OxCoder-9B, a model of the same depth, uses it in \layerOxTwentyFour{} percent. A fixed window of layers cannot follow such differences, but a selection per sample can.

\answer{2}{The carrier neurons should be selected by \semcon{}, which is forward-only and scores a neuron in the coordinates of the patch. This selection gives the highest Pass@1 in \attrCellsBest{} of \attrCellsTotal{} settings, significantly above both gradient-based attributions in \attrCellsSig{} of them and never significantly below. The patch then changes the carriers \ratioMin{} to \ratioMax{} times less in relative norm than those of the gradient-based attributions and turns their write vectors without rescaling them. It lies in late layers, never in the first \layerNeverPct{} percent of the depth.}

\subsection{Effectiveness and Efficiency}
\label{sec:rq3}

\begin{table}[htb]
\centering\tabsmall
\caption{Pass@1 in percent in the main comparison, on all 622 PyEvo and 708 RustEvo samples per model. Unrepaired is the model without a repair. The best value of each row is in bold, and higher values are better.}
\label{tab:main}
\begin{tabular}{llrrrrr}
\toprule
\textbf{Model} & \textbf{Benchmark} & \textbf{Unrepaired} & \textbf{AlphaEdit} & \textbf{STAR} & \textbf{LowRank} & \textbf{\astra{}} \\
\midrule
MiniCPM5-1B & PyEvo & 7.4 & 17.0 & 25.1 & 7.7 & \textbf{40.2} \\
 & RustEvo & 6.2 & 26.4 & 45.2 & 11.6 & \textbf{77.1} \\
\midrule
OpenCoder-8B & PyEvo & 17.4 & 31.2 & 39.7 & 23.5 & \textbf{61.9} \\
 & RustEvo & 37.1 & 57.8 & 64.0 & 52.4 & \textbf{90.8} \\
\midrule
OxCoder-9B & PyEvo & 20.3 & 37.1 & 44.1 & 45.2 & \textbf{47.3} \\
 & RustEvo & 32.5 & 62.9 & 72.5 & 74.3 & \textbf{82.8} \\
\bottomrule
\end{tabular}

\end{table}

On average over the \cellsAllTotal{} settings of the main comparison (Table~\ref{tab:main}), \astra{} reaches \avgAllAstra{} percent Pass@1, against \avgAllNone{} percent for the unrepaired models and \avgAllBest{} percent for \avgAllBestName{}, the baseline with the highest average. \astra{} is the best method in all \cellsAllTotal{} settings, and its difference from every baseline is significant in \cellsAllSig{} of them (paired McNemar test, $p<0.05$). The exception is OxCoder-9B with PyEvo, where it is on par with STAR and LowRank (\pOxPyAstra{} against \pOxPyStar{} and \pOxPyLow{} percent). Its gain over the best baseline of a setting ranges from \gainAllCellMin{} to \gainAllCellMax{} points.

Pooled over the \pooledAllN{} samples of the \cellsAllTotal{} settings, \astra{} solves \pooledAllWinStar{} samples that STAR does not and loses \pooledAllLossStar{} (\pooledAllPStar{}), and the corresponding counts are \pooledAllWinLow{} and \pooledAllLossLow{} against LowRank and \pooledAllWinAlpha{} and \pooledAllLossAlpha{} against AlphaEdit. Moreover, after a Holm correction over the \holmAllTotal{} comparisons with the three baselines in single settings, \holmAllSig{} remain significant, and the remaining \holmAllNot{} are those with STAR and LowRank on OxCoder-9B with PyEvo. The average gain over the best baseline (\avgAllBestName{}) is \gainAllBest{} points, with an interval of \gainAllLo{} to \gainAllHi{} points. It amounts to \gainSmBest{} points (\gainSmLo{} to \gainSmHi{}) on the small model and \gainBest{} points (\gainLo{} to \gainHi{}) on the large ones.

The baselines depend strongly on the model, whereas \astra{} is at least on par with the best baseline everywhere. LowRank hardly improves MiniCPM5-1B but is the best baseline on OxCoder-9B, whereas STAR is the best baseline on the other two models. The two gaps are consistent with this difference. AlphaEdit writes into layers 4 to 8, which in all three models lie in the depth that \semcon{} never selects. It targets one residual at a single position, whereas \astra{} chooses late layers per sample and corrects all failing tokens jointly.

\begin{table}[htb]
\centering\tabsmall
\caption{Pass@1 in percent by type of API change in the main comparison, on all samples of the \cellsAllTotal{} settings. PyEvo has \chNSignaturePy{}, \chNImplicitPy{}, \chNStabilizedPy{} and \chNDeprecatedPy{} samples of the four types and RustEvo has \chNSignatureRu{}, \chNImplicitRu{}, \chNStabilizedRu{} and \chNDeprecatedRu{}. Best is the best baseline for the setting and the type, the better value of each pair is in bold, and higher values are better.}
\label{tab:changecell}
\begin{tabular}{llrrrrrrrr}
\toprule
 &  & \multicolumn{2}{c}{\textbf{Signature change}} & \multicolumn{2}{c}{\textbf{Implicit change}} & \multicolumn{2}{c}{\textbf{Stabilized function}} & \multicolumn{2}{c}{\textbf{Deprecated API}} \\
\cmidrule(lr){3-4}\cmidrule(lr){5-6}\cmidrule(lr){7-8}\cmidrule(lr){9-10}
\textbf{Model} & \textbf{Benchmark} & \textbf{Best} & \textbf{\astra{}} & \textbf{Best} & \textbf{\astra{}} & \textbf{Best} & \textbf{\astra{}} & \textbf{Best} & \textbf{\astra{}} \\
\midrule
MiniCPM5-1B & PyEvo & 23.3 & \textbf{31.6} & 27.9 & \textbf{40.3} & 20.0 & \textbf{40.5} & 36.8 & \textbf{55.9} \\
 & RustEvo & 52.9 & \textbf{80.6} & 47.0 & \textbf{79.7} & 37.8 & \textbf{71.6} & 22.2 & \textbf{70.4} \\
\midrule
OpenCoder-8B & PyEvo & 32.3 & \textbf{48.1} & 41.8 & \textbf{62.7} & 36.4 & \textbf{66.4} & 58.8 & \textbf{72.1} \\
 & RustEvo & 66.9 & \textbf{92.1} & 76.0 & \textbf{93.1} & 52.7 & \textbf{87.8} & 33.3 & \textbf{85.2} \\
\midrule
OxCoder-9B & PyEvo & \textbf{41.4} & 40.6 & 46.3 & \textbf{49.8} & 44.1 & \textbf{44.5} & \textbf{61.8} & \textbf{61.8} \\
 & RustEvo & 79.3 & \textbf{86.0} & 77.4 & \textbf{83.9} & 73.0 & \textbf{79.7} & 51.9 & \textbf{70.4} \\
\bottomrule
\end{tabular}

\end{table}

Table~\ref{tab:changecell} shows that the gain is not confined to one type of API change. \astra{} is the best method in \chCellsWin{} of the \chCellsTotal{} combinations of a setting and a type, including all combinations of MiniCPM5-1B and of OpenCoder-8B. It ties with the best baseline for deprecated APIs on OxCoder-9B with PyEvo and is \chCellsMaxLoss{} points below it for signature changes in the same setting. When the samples of all settings are pooled, it is best for each of the four types, and its gain over the best baseline lies between \chAllGainMin{} and \chAllGainMax{} points. The gain is largest for \chAllGainMaxType{}, the type on which the unrepaired models are weakest (\chAllGainMaxNone{} percent, against \chAllRestNoneMin{} to \chAllRestNoneMax{} percent for the other types). A stabilized function was unstable or absent in earlier releases, so the result suggests that a patch can introduce an interface and not only redirect the model between two forms that it knows.

\begin{table}[htb]
\centering\tabsmall
\caption{Outcomes of the main comparison in percent of all samples of the \cellsAllTotal{} settings. Pass is an answer that passes all tests, Test an answer that fails a test, API an answer that does not use the required interface, Comp an answer that does not compile, and Other an unusable answer, one without the required function signature, without extractable code or with a timeout. The arrows mark the better direction.}
\label{tab:outcome}
\begin{tabular}{llrrrrrrrrr}
\toprule
 &  & \multicolumn{4}{c}{\textbf{PyEvo}} & \multicolumn{5}{c}{\textbf{RustEvo}} \\
\cmidrule(lr){3-6}\cmidrule(lr){7-11}
\textbf{Model} & \textbf{Method} & \textbf{Pass $\uparrow$} & \textbf{Test $\downarrow$} & \textbf{API $\downarrow$} & \textbf{Other $\downarrow$} & \textbf{Pass $\uparrow$} & \textbf{Comp $\downarrow$} & \textbf{API $\downarrow$} & \textbf{Test $\downarrow$} & \textbf{Other $\downarrow$} \\
\midrule
MiniCPM5-1B & Unrepaired & 7.4 & 74.3 & 15.4 & 2.9 & 6.2 & 55.2 & 36.3 & 2.1 & 0.1 \\
 & AlphaEdit & 17.0 & 70.3 & 8.4 & 4.3 & 26.4 & 44.8 & 23.3 & 2.5 & 3.0 \\
 & STAR & 25.1 & 54.8 & 11.3 & 8.8 & 45.2 & 38.7 & 10.9 & 1.8 & 3.4 \\
 & LowRank & 7.7 & 80.4 & 9.6 & 2.3 & 11.6 & 41.5 & 45.2 & 1.1 & 0.6 \\
 & \astra{} & 40.2 & 52.3 & 4.5 & 3.1 & 77.1 & 15.1 & 3.4 & 1.4 & 3.0 \\
\midrule
OpenCoder-8B & Unrepaired & 17.4 & 70.7 & 9.6 & 2.3 & 37.1 & 31.8 & 23.9 & 6.8 & 0.4 \\
 & AlphaEdit & 31.2 & 60.5 & 6.1 & 2.3 & 57.8 & 23.3 & 12.1 & 5.5 & 1.3 \\
 & STAR & 39.7 & 53.9 & 5.1 & 1.3 & 64.0 & 21.0 & 11.7 & 3.2 & 0.0 \\
 & LowRank & 23.5 & 68.8 & 6.6 & 1.1 & 52.4 & 32.1 & 9.6 & 5.6 & 0.3 \\
 & \astra{} & 61.9 & 33.8 & 2.7 & 1.6 & 90.8 & 5.9 & 1.8 & 1.1 & 0.3 \\
\midrule
OxCoder-9B & Unrepaired & 20.3 & 71.1 & 7.4 & 1.3 & 32.5 & 47.3 & 14.3 & 5.5 & 0.4 \\
 & AlphaEdit & 37.1 & 56.1 & 5.3 & 1.4 & 62.9 & 31.2 & 3.1 & 2.3 & 0.6 \\
 & STAR & 44.1 & 46.1 & 4.5 & 5.3 & 72.5 & 20.8 & 2.7 & 1.4 & 2.7 \\
 & LowRank & 45.2 & 50.3 & 3.2 & 1.3 & 74.3 & 20.2 & 3.0 & 2.0 & 0.6 \\
 & \astra{} & 47.3 & 48.2 & 3.2 & 1.3 & 82.8 & 13.8 & 1.3 & 1.3 & 0.8 \\
\bottomrule
\end{tabular}

\end{table}

The outcome classes of Table~\ref{tab:outcome} show how stale API knowledge manifests and what a repair changes. The unrepaired models rarely produce unusable answers. Instead, they write well-formed code for an interface that no longer exists or that behaves differently. On RustEvo, the compiler rejects such code (\ocAllRuNoneCompile{} percent of the samples over the three models) or the answer avoids the required interface (\ocAllRuNoneApi{} percent). On PyEvo, where an outdated interface surfaces only at run time, the dominant class is the failed test (\ocAllPyNoneTest{} percent). \astra{} lowers the two RustEvo shares to \ocAllRuAstraCompile{} and \ocAllRuAstraApi{} percent, whereas the best baseline for each outcome leaves \ocAllRuBestCompile{} and \ocAllRuBestApi{} percent. It has the lowest share of compilation failures in all \ocCompileCells{} RustEvo settings, and no baseline has a lower share of answers without the API in any of the \ocApiCells{} settings. On PyEvo, it lowers failed tests to \ocAllPyAstraTest{} percent.

The gains thus come from the classes that the outdated interface causes, while the share of unusable answers of \astra{} stays at or below \ocOtherMaxAstra{} percent. The patch changes which interface the model writes and leaves the form of its answers intact. After a repair by \astra{}, failed tests remain the largest class of failures on PyEvo. A failed test does not separate a wrong interface from an error in the remaining program logic, so a part of these failures may lie outside what a repair of interface knowledge can reach. This would also account for OxCoder-9B with PyEvo, where STAR, LowRank and \astra{} end at similar shares of failed tests and of Pass@1.

\begin{figure}[htb]
\centering
\begin{tikzpicture}
\begin{axis}[paperfig, paperbars, bar width=4.6pt, ymin=0, ymax=10,
  ytick={0,2,4,6,8,10}, ylabel={Repair time (seconds)},
  xtick={1,2,3,4,5,6}, xticklabels={PyEvo,RustEvo,PyEvo,RustEvo,PyEvo,RustEvo}, xmin=0.4, xmax=6.6, after end axis/.code={\node[font=\scriptsize, anchor=north] at (rel axis cs:0.1774,-0.15) {MiniCPM5-1B};\node[font=\scriptsize, anchor=north] at (rel axis cs:0.5000,-0.15) {OpenCoder-8B};\node[font=\scriptsize, anchor=north] at (rel axis cs:0.8226,-0.15) {OxCoder-9B};}]
\addplot[fill=mAlpha, draw=black!60] coordinates {(1,2.66) (2,2.57) (3,6.80) (4,5.64) (5,7.65) (6,6.80)};
\addplot[fill=mStar, draw=black!60] coordinates {(1,4.53) (2,3.63) (3,7.17) (4,5.69) (5,7.33) (6,6.43)};
\addplot[fill=mLow, draw=black!60] coordinates {(1,0.75) (2,0.70) (3,1.46) (4,1.14) (5,1.98) (6,1.65)};
\addplot[fill=mAstra, draw=black!60] coordinates {(1,1.68) (2,1.60) (3,3.01) (4,2.64) (5,4.09) (6,3.05)};
\legend{AlphaEdit, STAR, LowRank, \astra{}}
\end{axis}
\end{tikzpicture}
\caption{Mean repair time per sample in seconds in the main comparison, without the generation of the answer.}
\label{fig:time}
\end{figure}
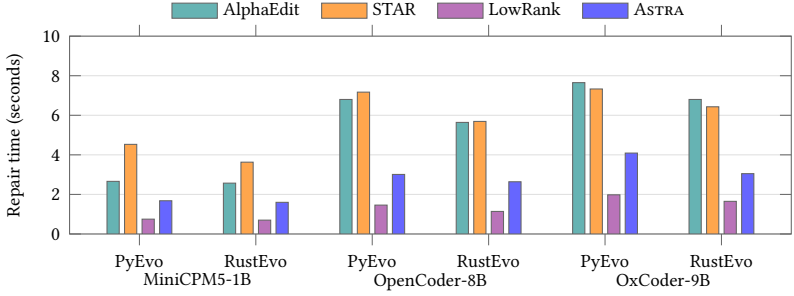

Regarding efficiency, Figure~\ref{fig:time} shows the mean repair time per sample. \astra{} needs between \timeAllCellMinAstra{} and \timeAllCellMaxAstra{} seconds in the single settings and is faster than STAR and AlphaEdit in all \timeAllCellsFasterStar{} of them, whereas LowRank is the fastest method everywhere. On average over the \cellsAllTotal{} settings, \astra{} needs \timeAllAstra{} seconds, against \timeAllLow{} for LowRank, \timeAllStar{} for STAR and \timeAllAlpha{} for AlphaEdit. It is thus \timeAllRatioStar{} times faster than STAR and \timeAllRatioAlpha{} times faster than AlphaEdit and takes about twice as long as the ten steps of LowRank. The times were measured while several processes shared a GPU, so their ratios are the meaningful quantity. Forward passes dominate the cost: each round needs one forward pass per view and one small linear solve, and no step needs a gradient.

\begin{table}[htb]
\centering\tabsmall
\caption{Ablation study of \astra{}: Pass@1 in percent of \astra{} and of variants that change one component, on the shared subset of all \cellsAllTotal{} settings. Sequential corrects the failing tokens one after another, Margin 1 replaces the margin of five logits by one logit, One view fits only the original repair prompt, and Two views uses two views and eight neurons per answer token. The best value of each row is in bold, and higher values are better.}
\label{tab:ablation}
\begin{tabular}{llrrrrr}
\toprule
\textbf{Model} & \textbf{Benchmark} & \textbf{\astra{}} & \textbf{Sequential} & \textbf{Margin 1} & \textbf{One view} & \textbf{Two views} \\
\midrule
MiniCPM5-1B & PyEvo & \textbf{37.0} & 9.1 & 8.2 & 28.8 & 32.7 \\
 & RustEvo & \textbf{74.6} & 27.5 & 23.7 & 66.1 & 69.5 \\
\midrule
OpenCoder-8B & PyEvo & \textbf{61.1} & 26.0 & 32.2 & 56.2 & 56.7 \\
 & RustEvo & \textbf{89.0} & 56.4 & 56.4 & 87.3 & 85.2 \\
\midrule
OxCoder-9B & PyEvo & \textbf{46.2} & 18.3 & 27.9 & 37.0 & 43.3 \\
 & RustEvo & \textbf{80.5} & 47.5 & 53.8 & 73.7 & 75.0 \\
\bottomrule
\end{tabular}

\end{table}

Finally, the ablation study of Table~\ref{tab:ablation} shows which components of \astra{} produce this effectiveness. All statements of significance on the variants refer to a paired McNemar test against \astra{} with $p<0.05$. The sequential solve lowers Pass@1 by \ablAllDSeq{} points when the samples of all settings are pooled and by \ablAllDMinSeq{} to \ablAllDMaxSeq{} points in the single settings, significantly in each of the \ablAllSigSeq{} settings. This is the largest loss of all variants. It measures on the repair side what the tailoring gap costs, and the linear system explains it. The failing tokens of a sample share their carriers, so the correction of one token moves the leads of the others through the off-diagonal entries of $A$. With \rqTokMin{} to \rqTokMax{} failing tokens per sample on average, a token-by-token procedure repeatedly undoes its earlier corrections. Similarly, a margin of one logit instead of five lowers Pass@1 by \ablAllDMargin{} points, significantly in every setting. The patch is computed under teacher forcing on the views of the repair prompt, whereas the test generates the answer freely from another prompt, and the loss indicates that a lead of one logit does not survive this change.

Among the remaining variants, fitting one view lowers Pass@1 by \ablAllDOne{} points and is below the full method in all settings, significantly in \ablAllSigOne{} of them. The variant with two views and half the budget is \ablAllDOld{} points lower and also below the full method everywhere, significantly in \ablAllSigOld{} settings. The views are thus an effective enhancement. They are renderings of the same repair prompt and add no information, yet fitting them together gains \ablAllDOne{} points. Even with one view, \astra{} reaches \oneViewAll{} percent Pass@1 over all settings, against \oneViewBestAll{} percent for \oneViewBestName{}. The joint solve and the margin decide whether a repair holds, and more views and a larger budget add several points.

\answer{3}{\astra{} is the most effective method in all \cellsAllTotal{} settings and for every type of API change, with \avgAllAstra{} percent Pass@1 on average, \gainAllBest{} points above the best baseline (\avgAllBestName{}, interval \gainAllLo{} to \gainAllHi{} points). It repairs a sample in \timeAllAstra{} seconds on average, which is \timeAllRatioStar{} times faster than STAR and \timeAllRatioAlpha{} times faster than AlphaEdit, but takes about twice as long as LowRank. Of its components, the joint solve and the margin matter most, since their removal costs \ablAllDSeq{} and \ablAllDMargin{} points, whereas fewer views and a smaller budget cost \ablAllDOne{} and \ablAllDOld{} points.}

\subsection{Side Effects and Robustness}
\label{sec:rq4}

\begin{table}[htb]
\centering\tabsmall
\caption{Side effects on HumanEval after the repair of one problem, averaged over \ovRepairs{} repairs per model and method. Acc-APD and EM-APD are the absolute changes of the teacher-forced token accuracy and of exact match in points, and BLEU-RPD is the relative change of BLEU in percent, each measured on the \ovHeldOut{} held-out problems against the unrepaired model, so a negative value is a loss. The arrows mark the better direction.}
\label{tab:overfit}
\begin{tabular}{llrrr}
\toprule
\textbf{Model} & \textbf{Method} & \textbf{Acc-APD $\uparrow$} & \textbf{EM-APD $\uparrow$} & \textbf{BLEU-RPD $\uparrow$} \\
\midrule
MiniCPM5-1B & AlphaEdit & $-$3.26 & +1.35 & $-$10.77 \\
 & STAR & $-$0.20 & +0.36 & $-$1.66 \\
 & LowRank & +0.54 & +0.38 & $-$0.15 \\
 & \astra{} & $-$0.69 & +0.71 & $-$4.11 \\
\midrule
OpenCoder-8B & AlphaEdit & $-$0.36 & +0.76 & $-$1.42 \\
 & STAR & +0.43 & +0.00 & +0.16 \\
 & LowRank & +0.61 & +0.02 & +0.09 \\
 & \astra{} & +0.45 & +0.07 & $-$0.14 \\
\midrule
OxCoder-9B & AlphaEdit & $-$0.44 & +1.66 & $-$4.83 \\
 & STAR & +0.62 & +1.99 & $-$1.59 \\
 & LowRank & +0.29 & +0.59 & $-$2.37 \\
 & \astra{} & +0.81 & +1.09 & $-$2.03 \\
\bottomrule
\end{tabular}

\end{table}

A repair should not disturb unrelated behavior. For repairs of single HumanEval problems, Table~\ref{tab:overfit} reports the overfitting measures that the evaluation of STAR introduced~\cite{gu2025star}. On the large models, the changes of \astra{} are small. Its Acc-APD is \ovAccOpAstra{} and \ovAccOxAstra{} points, its EM-APD is not negative, and its BLEU-RPD is \ovBleuOpAstra{} and \ovBleuOxAstra{} percent. This is comparable to STAR and LowRank (at most \ovBleuMaxOther{} percent in absolute value) and smaller than for AlphaEdit (\ovBleuOpAlpha{} and \ovBleuOxAlpha{} percent). On the small model, however, the side effects of AlphaEdit and \astra{} are larger, and only LowRank stays below 1 percent of BLEU-RPD. There, the BLEU-RPD of \astra{} (\ovBleuMoAstra{} percent) lies between those of STAR and AlphaEdit (\ovBleuMoStar{} and \ovBleuMoAlpha{} percent), and its Acc-APD is \ovAccMoAstra{} points.

The footprint of the patch is consistent with the larger side effects on the small model. It has fewer feed-forward neurons, so the same budget of 16 neurons per answer token changes a larger share of them, between \neuronShareSmMin{} and \neuronShareSmMax{} percent against \neuronShareMin{} to \neuronShareMax{} percent on the large models, measured on the API repair samples. The relative patch norm, by contrast, is not larger on MiniCPM5-1B than on the large models. AlphaEdit, STAR and \astra{}, which write directly into existing weights, all show their largest BLEU-RPD on the small model, which suggests that a small model has less spare capacity. Moreover, the configuration was chosen with the large models.

\begin{table}[htb]
\centering\tabsmall
\caption{Pass@1 in percent of all methods under the original test prompt (Orig) and under a specification prompt (Spec), a phrasing that resembles none of the views, on the shared subset of all \cellsAllTotal{} settings. The best value of each row and phrasing is in bold, and higher values are better.}
\label{tab:rob}
\begin{tabular}{llrrrrrrrrrr}
\toprule
 &  & \multicolumn{2}{c}{\textbf{Unrepaired}} & \multicolumn{2}{c}{\textbf{AlphaEdit}} & \multicolumn{2}{c}{\textbf{STAR}} & \multicolumn{2}{c}{\textbf{LowRank}} & \multicolumn{2}{c}{\textbf{\astra{}}} \\
\cmidrule(lr){3-4}\cmidrule(lr){5-6}\cmidrule(lr){7-8}\cmidrule(lr){9-10}\cmidrule(lr){11-12}
\textbf{Model} & \textbf{Benchmark} & \textbf{Orig} & \textbf{Spec} & \textbf{Orig} & \textbf{Spec} & \textbf{Orig} & \textbf{Spec} & \textbf{Orig} & \textbf{Spec} & \textbf{Orig} & \textbf{Spec} \\
\midrule
MiniCPM5-1B & PyEvo & 7.2 & 8.7 & 15.9 & 22.1 & 22.1 & 27.4 & 5.3 & 13.9 & \textbf{37.0} & \textbf{55.8} \\
 & RustEvo & 5.1 & 3.8 & 25.4 & 33.1 & 41.9 & 46.6 & 12.3 & 16.9 & \textbf{74.6} & \textbf{85.2} \\
\midrule
OpenCoder-8B & PyEvo & 13.5 & 12.0 & 28.8 & 27.4 & 35.6 & 37.0 & 19.2 & 18.3 & \textbf{61.1} & \textbf{59.6} \\
 & RustEvo & 34.7 & 23.7 & 53.8 & 40.3 & 62.3 & 51.3 & 47.5 & 44.9 & \textbf{89.0} & \textbf{85.6} \\
\midrule
OxCoder-9B & PyEvo & 15.4 & 15.9 & 35.6 & 35.1 & 40.9 & 43.3 & 42.3 & 51.4 & \textbf{46.2} & \textbf{52.9} \\
 & RustEvo & 28.4 & 17.8 & 63.6 & 55.1 & 72.9 & 70.8 & 79.7 & 79.7 & \textbf{80.5} & \textbf{86.4} \\
\bottomrule
\end{tabular}

\end{table}

For robustness, we replace the test prompt, which has the wording of the original task, by an unseen phrasing: a specification prompt, a terse list that states the goal, the required API with its module and signature, and the instruction to output only the code block. \astra{} keeps its advantage under this phrasing (Table~\ref{tab:rob}). It is ahead of the baselines and the unrepaired model in all \robCellsAheadAll{} settings, and its average gain over the best baseline of a setting is \robGapSpecBest{} points, against \robGapOrigBest{} under the original prompt. Its own Pass@1 decreases only on OpenCoder-8B, by at most \robMaxDropAstra{} points.

The columns of the unrepaired model show that the new phrasing is not easier by itself. Although the specification prompt names the required API, the average Pass@1 of the unrepaired models falls from \robAllOrigNone{} to \robAllSpecNone{} percent, whereas that of \astra{} rises from \robAllOrigAstra{} to \robAllSpecAstra{} percent. LowRank rises from \robAllOrigLow{} to \robAllSpecLow{} percent, and AlphaEdit and STAR stay at \robAllSpecAlpha{} and \robAllSpecStar{} percent, close to their values under the original prompt. A patch that had memorized the wording of the repair prompt would lose under a phrasing that it has not seen. The result instead agrees with the purpose of the views and of the margin, which fit the patch to the interface under several wordings and leave a lead that survives a further one, and with the loss of the single-view variant in the ablation.

\answer{4}{A repair by \astra{} changes unrelated code generation little on the large models, where the BLEU-RPD is \ovBleuOpAstra{} and \ovBleuOxAstra{} percent and comparable to STAR and LowRank, and more on the small model (\ovBleuMoAstra{} percent). Its advantage is robust to an unseen phrasing of the test prompt, under which it remains ahead of all three baselines and of the unrepaired model in all \robCellsAheadAll{} settings, with an average gain of \robGapSpecBest{} points over the best baseline.}

\section{Discussion}
\label{sec:discussion}

Repair methods are usually assembled from separate stages: an attribution that ranks components, a selection of the highest-ranked ones, and an update rule. The two gaps show, however, that these stages must be designed together. A stronger attribution is of little use if the update ignores which failure is being repaired, and the largest losses in the ablation of \astra{} come from the update, namely from a token-by-token solve and an insufficient margin. Three design choices of \astra{} follow from this view. First, the attribution is expressed in the coordinates of the update, so a high score identifies a neuron that moves the failing contrast under a small change. Second, the patch is solved jointly in closed form, because the corrections of different tokens interact through shared carrier neurons. Third, several views are fitted together, because a patch fitted to one phrasing does not hold under another. Moreover, agreement with a reference set of neurons is not required for a repair, since many sets of carriers satisfy the same requirements. The size of the patch that the selected carriers need is more informative: where the gradient-based attributions demand the largest patches, they also repair significantly worse. The relative patch norm is thus an inexpensive diagnostic that is available at repair time, before any test is run. These findings reach beyond API evolution. The two gaps give neuron-level repair a way to assess its own pipeline: as in fault localization for programs, an attribution is judged by the repair that it enables and not by its ranking alone. \astra{} itself needs only forward passes and the output embedding of the model, and one configuration served three models of different size and architecture, which suggests that it carries over to other decoder-only models and to other failures that show in the tokens of an answer. For the maintenance of code models, this makes a repair comparable to a hotfix between two retrainings. A repair that takes seconds and needs no training loop can be run when a failure is reported, and it reaches every tool that uses the model. The patch changes a small set of columns in a few late layers, so it can be stored as a small difference of the model, reviewed, traced to its failure, and reverted. Nevertheless, a regression check remains necessary before a patch is accepted, since the side effects are small on the large models and larger on the small one.

\section{Conclusion}
\label{sec:conclusion}

In this paper, we studied the repair of code language models whose knowledge of library interfaces is outdated. We examined the two assumptions behind existing neuron-level repair and identified two coupled gaps: the targeting gap between the neurons that failure attribution targets and the neurons that can carry a patch, and the tailoring gap between a generic update and a patch built for the failure. To address them, we proposed \astra{}, which targets carrier neurons by \semcon{} and tailors a closed-form patch that corrects all failing tokens jointly. Experiments on two benchmarks and three code models show that \astra{} is the most effective method in every setting and for every type of API change, and that it repairs a sample within seconds. Its advantage persists under an unseen phrasing of the test prompt, and its side effects are comparable to those of STAR and LowRank on the two large models and more pronounced on the small one. Our analysis further shows that \semcon{} selects carrier neurons that need a smaller patch than those of gradient-based attributions. These results suggest that attribution and update should be designed together when a language model is repaired.

\bibliographystyle{ACM-Reference-Format}
\bibliography{refs}

\end{document}